\documentclass[sigplan,10pt]{acmart}

\renewcommand\footnotetextcopyrightpermission[1]{}
\usepackage[utf8]{inputenc} %
\usepackage[T1]{fontenc}    %
\usepackage{hyperref}       %
\usepackage{url}            %
\usepackage{amsfonts}       %
\usepackage{nicefrac}       %
\usepackage{xcolor}         %
\usepackage{wrapfig}
\usepackage{caption, subcaption}
\usepackage{kotex}
\usepackage{xspace}

\usepackage{booktabs} %
\usepackage{multirow}
\usepackage{colortbl}
\usepackage{makecell}

\usepackage{graphicx}              %
\graphicspath{ {figs/} {graphs/} }
\usepackage[inkscapelatex=false]{svg}

\usepackage{pifont}
\usepackage{tikz}
\newcommand*\circled[1]{\tikz[baseline=(char.base)]{
            \node[shape=circle,fill,inner sep=0.3pt] (char) {\textcolor{white}{#1}};}}

\usepackage{setspace}
\usepackage[noend]{algpseudocode}
\usepackage{algorithm}

\makeatletter
\def\BState{\State\hskip-\ALG@thistlm}
\makeatother
\algnewcommand{\LeftComment}[1]{\State // \small \textit{#1}}
\algnewcommand{\InlineComment}[1]{// \small \textit{#1}}

\RequirePackage[normalem]{ulem}
\RequirePackage{color}\definecolor{RED}{rgb}{1,0,0}\definecolor{BLUE}{rgb}{0,0,1}

\newcommand{\sys}{\textsc{SpecScale}\xspace}

\begin{document}

\title{Taming Speculative Search for Test-Time Scaling in LLM Serving}
\author{Jinwoo Jeong}
\affiliation{%
  \institution{Korea University}
  \city{Seoul}
  \country{Republic of Korea}}
\email{jwjeong@csl.korea.ac.kr}

\author{Woohyung Choi}
\affiliation{%
  \institution{Korea University}
  \city{Seoul}
  \country{Republic of Korea}}
\email{whchoi@csl.korea.ac.kr}

\author{Myeongjae Jeon}
\affiliation{%
  \institution{POSTECH}
  \city{Pohang}
  \country{Republic of Korea}}
\email{mj.jeon@postech.ac.kr}

\author{Jeongseob Ahn}
\affiliation{%
  \institution{Korea University}
  \city{Seoul}
  \country{Republic of Korea}}
\email{jsahn@csl.korea.ac.kr}

\begin{abstract}
Test-time scaling has recently emerged as a powerful approach for improving LLM reasoning
by allocating additional computation during inference, substantially enhancing accuracy
on challenging tasks such as mathematics and coding. 
To accelerate the exploration of reasoning paths, recent studies proposed speculative execution. 
However, we show that supporting speculative execution poses two unique challenges for LLM serving systems:
(1) an explosion in the search space of candidate paths and (2) frequent, fine-grained verification tasks
for candidates.

To address these challenges, this paper proposes {\it \sys}, a serving system for efficient
speculative execution. We introduce three techniques to reconcile the trade-off between latency and
computational overhead: (1) early pruning of low-quality candidate paths,
(2) deduplicating computation across redundant candidate paths, and (3) deferring fine-grained
verification tasks.
We evaluate \sys on challenging reasoning benchmarks, including MATH and Olympiad. Our results show that
\sys significantly outperforms both non-speculative and recent speculative approaches, delivering 
substantial improvements in throughput and latency while preserving answer quality.
\end{abstract}

\maketitle %
\pagestyle{plain}\thispagestyle{plain} %

\section{Introduction}
In the last few years, the success of large language models (LLMs) has been largely driven by
training-time scaling: increasing model parameters, training data, and compute budgets to improve
model quality~\cite{hoffmann:2022, kaplan:2020, hurst:2024, achiam:2024}.
This paradigm has delivered remarkable progress across a wide range of tasks,
from natural language understanding to code generation. However, continued scaling along this
axis faces fundamental limitations. Training state-of-the-art models now demands massive 
computational resources, making further scaling increasingly costly and inaccessible~\cite{gholami:2024, kaplan:2020, hoffmann:2022}.

Recent work has highlighted an orthogonal direction: {\it test-time scaling}~\cite{snell:2025, muennighoff:2025,
shunyu:2023, madaan:2023, brown:2024, wang:2024}.
Instead of training ever-larger models, test-time scaling improves model performance by allocating more
computation for smaller models during inference. At a high level, test-time scaling generates multiple
candidates (reasoning paths) for a single query, and then uses a separate learned
reward model~\cite{lightman:2024, qwen-prm} to verify these candidates and select the
most promising ones. The search then continues only along the selected candidates, repeating 
this ``{\it expand--verify--select}'' loop (called a step) until at least one complete solution
is found. Notably, these methods have demonstrated substantial gains on complex reasoning 
tasks such as mathematical problem solving without modifying model parameters or
retraining~\cite{snell:2025, wu:2025, beeching:2024}.

While test-time scaling is promising, its search process can incur substantial idle time
because multiple candidate paths expand in parallel with diverging lengths, and each step
requires verification before the search can proceed. 
This synchronous scheduling inevitably leaves faster candidates
idle while slower ones catch up, causing significant waiting time at every step.
To reduce the idle time, speculative execution has emerged
as a key mechanism~\cite{sixu:2025, chen:2025}. Instead of waiting for all candidates to be
verified at each step, it speculatively expands any candidate as soon as it
finishes its current step, anticipating that it will remain among the top-$k$ candidates. This strategy
can substantially reduce waiting time and improve end-to-end latency, since promising candidates
continue to make progress while others catch up. 

However, we observe that speculative execution of reasoning paths constitutes a fundamentally
different workload for LLM serving systems compared to conventional inference. It exhibits two
distinct characteristics that existing systems are not designed to handle. First, speculation
causes an explosion in the candidate space, especially in batched inference settings typical of
online LLM serving. As candidates can be expanded before their scores are known, the speculative
approach explores many more candidate paths than the non-speculative approach. This amplifies
decoding computation and increases KV memory pressure.
Second, verification is triggered independently for each candidate path upon completion, leading
to frequent, fine-grained verification tasks. These small and irregular tasks not only underutilize
the GPU, but also repeatedly interrupt the decode steps. Consequently, without appropriate
system support, the cost of speculation can outweigh the performance gains.

To address these challenges, we present \textbf{\sys}, a serving system designed to
efficiently support speculative execution through three key techniques. First,
we propose an \textit{\textbf{early pruning}} technique to proactively discard unpromising
paths before they consume excessive resources. Intuitively, not all speculative executions
are equally promising. Once a candidate’s score falls below the current top-$k$, it
can no longer enter the final top-$k$ set, rendering its continued speculative execution
unnecessary. Pruning these paths early significantly reduces the aggregate compute load
without compromising the accuracy of the final output. Furthermore, we can terminate the
search once $k$ perfect-score candidates have been observed, pruning all remaining explorations
because the top-$k$ set is already determined.

Second, to further reduce the computational overhead of exploring multiple candidate paths, we
introduce a \textbf{\textit{computation deduplication}} technique that shares decoding computation
across candidates with identical prefixes. In test-time scaling, our analysis identifies
that many candidate paths follow the same partial reasoning for several steps, creating 
opportunities to reuse computation on common prefixes. Instead of decoding each candidate
separately, \sys performs a single decode step for each distinct prefix and reuses the
resulting logits to advance all candidates that share that prefix. It only splits them into 
separate paths when their sampled next tokens diverge. 
Note that this technique can also be applied to the non-speculative approach.

Last, to address the overhead of increased verification tasks in speculation, we introduce \textbf{\textit{lazy verification}},
a deferred batching strategy. Instead of verifying every step immediately, verification requests
are enqueued and processed in large, GPU-efficient batches. However, deferring verification introduces
an inherent trade-off between latency and throughput: waiting longer yields larger, more efficient batches but 
can diminish the benefit of speculative execution. We navigate this trade-off using three policies
based on the amount of deferred work, sibling-group completion, and a per-sample deferral limit that
jointly determine when verification is triggered. This design maximizes hardware utilization
while preserving the latency gains of speculation.

We implement {\it \sys} as a lightweight LLM serving framework that supports continuous
batching~\cite{yu:2022}, paged attention~\cite{kwon:2023}, prefix KV
caching and sharing~\cite{zheng:2024}, and fused attention kernels~\cite{ye:2025}. We evaluate
four generator--verifier pairs drawn from the Qwen2.5 and Llama3 families on math reasoning
benchmarks, including GSM8K~\cite{gsm}, MATH-500~\cite{math}, and OlympiadBench~\cite{olympiad}.
All experiments are conducted on a single NVIDIA A100 GPU. Across all datasets, \sys
consistently outperforms both the naive-speculative and a recent speculation approach, FastTTS~\cite{chen:2025},
delivering substantial improvements in throughput and end-to-end latency while preserving
answer quality. On the MATH dataset, for the Qwen2.5-7B-Instruct paired with the
Qwen2.5-Math-PRM-7B, \sys achieves 2.18$\times$ and 1.87$\times$ higher throughput than the
naive speculation and FastTTS, respectively.

\section{Background}

\subsection{LLM Inference and Serving}
Transformer-based generative models consist of stacked self-attention and feed-forward
layers that compute contextual representations for input tokens, followed by a language
modeling (LM) head. The LM head produces logits, unnormalized scores over the vocabulary,
which are converted into a probability distribution through a Softmax function. 

Inference proceeds in two phases: {\it prefill} and {\it decode}.
In the prefill phase, the model processes the entire input token sequence in parallel to
generate the first output token. Then, in the decode phase, the model generates subsequent
tokens auto-regressively until an End-of-Sentence (EOS) token is produced or the maximum
generation length is reached~\cite{vaswani:2017}. 

During generation, the next output token is selected by sampling from the probability
distribution computed from the logits. The sampling process is influenced by a {\it temperature}
parameter, which scales the logits before the Softmax, thereby controlling the randomness of the next token~\cite{hinton2015distillingknowledgeneuralnetwork}.
A higher temperature increases the spread of the distribution, allowing the model to
generate more diverse tokens, while a lower temperature sharpens the distribution, making
the model more likely to output high-probability, deterministic tokens. Thus, the temperature
directly controls the trade-off between the randomness and coherence required by
the specific generative AI application.

To serve these models efficiently, modern LLM serving systems build on several foundational techniques
~\cite{yu:2022, agrawal:2024, kwon:2023, zheng:2024, dao:2022, ye:2025}.
However, they do not address the distinct
resource management challenges that arise specifically from test-time scaling methods.

\begin{figure}[t]
\centering
\includegraphics[width=3.in]{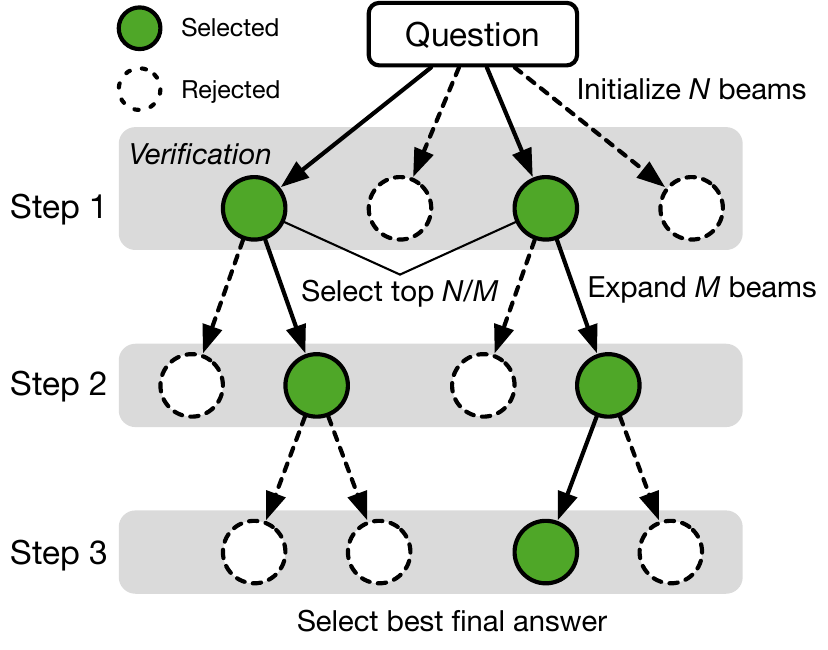}
\vspace{-0.10in}
\caption{Beam search with Process Reward Model (PRM), where beam size ($N$) is 4 and beam width ($M$) is 2}
\vspace{-0.10in}
\label{fig:beam_search}
\end{figure}

\subsection{Test-Time Scaling}

Test-time scaling improves accuracy by allocating additional computation during
inference~\cite{snell:2025, brown:2024, wang:2024}, yielding substantial gains on complex
reasoning tasks such as mathematics and coding~\cite{muennighoff:2025, dacheng:2025}.
Test-time scaling methods can be broadly categorized into two classes:
(1) internal, which encourages models to {\it think} by producing a long
Chain-of-Thought reasoning~\cite{openai:2024, deepseekai:2025}, and
(2) external, which improves reasoning performance using sampling or
verifier-guided methods with reward models~\cite{wu:2025, snell:2025}.
Our work focuses on the external approach, where a smaller model can achieve accuracy
comparable to that of a larger model by exploring and verifying multiple candidate
paths at inference time~\cite{beeching:2024}. 

In the external approach, a straightforward method is {\it best-of-$N$} sampling~\cite{touvron:2023},
where the LLM generates $N$ independent responses and an Outcome Reward Model (ORM)~\cite{uesato:2022}
selects the highest-scoring one. However, best-of-$N$ has intrinsic limitations in accuracy, because
it evaluates only final responses and does not guide intermediate reasoning.

\begin{table}[t]
\footnotesize
\centering
\begin{tabular}{c|c|c|c}
\toprule
Dataset         & Qwen2.5-3B  & Qwen2.5-32B    & \cellcolor[gray]{0.8} \makecell{Qwen2.5-3B\\w/ PRM (8 beams)}      \\
\midrule
GSM8K           & 88.90 \%    & 93.90 \%       & \cellcolor[gray]{0.8} 94.70 \%                                   \\
MATH-500        & 60.60 \%    & 80.30 \%       & \cellcolor[gray]{0.8} 76.24 \%                                   \\
OlympiadBench   & 19.52 \%    & 32.65 \%       & \cellcolor[gray]{0.8} 30.40 \%                                   \\
\bottomrule
\end{tabular}
\vspace{0.10in}
\caption{Effectiveness of test-time scaling on the math datasets in terms of accuracy}
\vspace{-0.20in}
\label{tab:accuracy}
\end{table}

To overcome these limitations, recent test-time scaling techniques, such as {\it beam search}~\cite{snell:2025}
and {\it Monte Carlo Tree Search} (MCTS)~\cite{lightman:2024, uesato:2022}, incrementally explore
and prune candidate paths, allowing step-wise guidance based on intermediate verification. Since MCTS
is known to incur substantially higher per-iteration cost due to repeated rollouts~\cite{chen:2025},
it is less practical for latency-sensitive online serving. In this work, we therefore focus
on beam search. 

\autoref{fig:beam_search} illustrates how test-time scaling with beam search can improve
the quality of model responses by incrementally exploring the solution space.
At each step, beam search keeps track of $N$ candidate sequences, where $N$ is called the
beam size. For a given input, the LLM initially generates $N$ independent sequences. Once
all $N$ candidate steps are complete, they are evaluated by a Process Reward Model
(PRM)~\cite{uesato:2022, lightman:2024}, which provides step-wise feedback by scoring
intermediate reasoning steps, and only the top-$k$ sequences with the highest scores are
retained for further expansion. Each selected sequence is then expanded into $M$ new 
candidates, where $M$ denotes the beam width.
To ensure that the number of active beams remains constant, the selection ratio is
configured such that top-$k = N/M$. This procedure effectively resembles a breadth-first
search that limits the search space while preserving high-quality trajectories. The process
continues until $N$ complete sequences are generated, and the final answer is selected based
on the highest weighted sum of the step-wise scores.

\autoref{tab:accuracy} shows that the smaller Qwen2.5-3B model with PRM-based test-time
scaling achieves accuracy comparable to the larger Qwen2.5-32B model on three math datasets.

\subsection{Inefficiency of Test-Time Scaling}

Although tree-based test-time scaling methods show strong performance in reasoning tasks~\cite{snell:2025, shunyu:2023,
ziyu:2024}, they introduce structural inefficiencies, primarily due to {\it waiting (or idle) time}.
Since the number of generated tokens varies across candidate sequences, the naive execution must wait
for all candidates at each step to complete verification before selecting the top-$k$ for the next step.
Even though some candidates complete early (e.g., their paths generate fewer tokens than others),
they cannot proceed to the next step until the scores of all concurrent candidates have been computed.
Consequently, this results in substantial waiting time at every step.

\begin{figure}[t]
\centering
\includegraphics[width=3.2in]{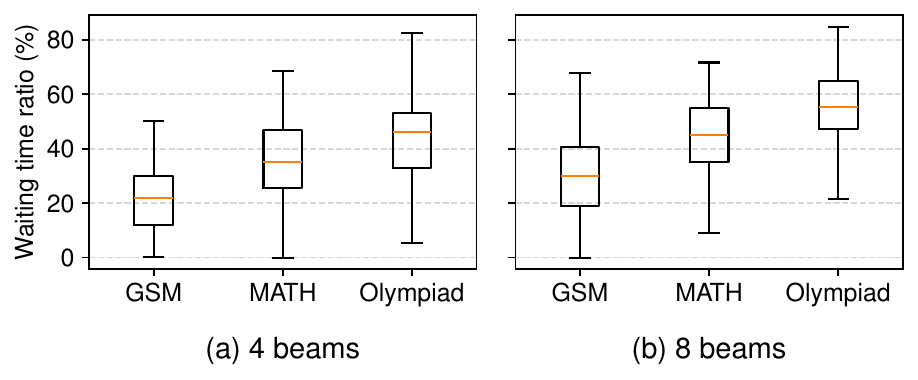}
\vspace{-0.10in}
\caption{Distribution of waiting time ratio for test-time scaling with beam search}
\vspace{-0.10in}
\label{fig:motiv_wait_time}
\end{figure}

\begin{figure*}[t]
\centering
\includegraphics[width=6.4in]{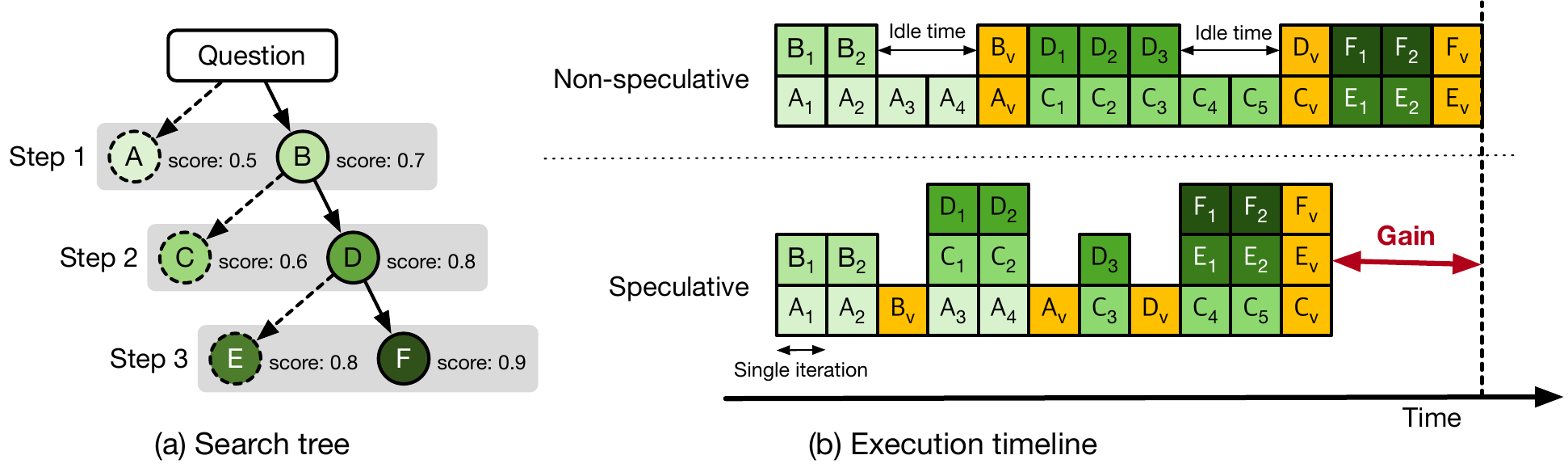}
\vspace{-0.10in}
\caption{Test-time scaling with two approaches: (a) example search tree, (b) execution timeline of
non-speculation and speculation. The numeric subscript on each rectangle indicates the order
of the decoding phases, while the subscript {\tt \textbf{v}} denotes verification.}
\label{fig:spec_approach}
\end{figure*}

We evaluate the waiting time under a single batch setting (batch size 1) on three math reasoning
datasets: GSM~\cite{gsm}, MATH~\cite{math}, and Olympiad~\cite{olympiad}.  Their difficulty increases
in the order of GSM (grade-school math problems), MATH (competition-level), and Olympiad.
\autoref{fig:motiv_wait_time} shows the distribution of waiting time ratios for the
{\tt Qwen2.5-3B-Instruct} model under test-time scaling with beam search. For each dataset,
the box plots are computed over 100 randomly sampled problems for beam sizes 4 and 8, respectively.
We use {\tt Qwen2.5-Math-PRM-7B} for verification. The median waiting time accounts for approximately
20--55\% of the total execution time, and this proportion becomes larger as
the problem difficulty increases. Thus, reducing waiting time is essential for achieving
efficient test-time scaling.

\section{Speculative Execution and Its Challenges}

\subsection{Speculative Execution}

Rather than waiting for all candidates to complete each step, speculative execution
expands any candidate that finishes early, anticipating that it is likely to be
selected for the next step~\cite{sixu:2025, chen:2025}. \autoref{fig:spec_approach} illustrates 
(a) an example search tree and (b) how speculation accelerates the naive process.
At Step 1, the {\tt A} and {\tt B} paths begin
generating tokens in parallel. In the non-speculative approach, although {\tt B} finishes
earlier than {\tt A} (e.g., {\tt B} produces two tokens while {\tt A}
produces four), it must wait until {\tt A} completes; only then are both candidates verified
together. In contrast, the speculative approach immediately expands {\tt B} to the next
step as soon as it finishes, without waiting for {\tt A}. When {\tt A}’s verification
later completes, if {\tt B} ultimately has a higher score than {\tt A}, the speculative
path has already made progress, effectively reducing execution time by eliminating unnecessary waiting.

\subsection{Challenges for Speculative Execution in Serving}
\label{sec:chall_spec}

Although speculation enables each path to progress independently without a synchronized
verification step, it introduces two fundamental challenges. First, speculation expands the search
space, substantially increasing the number of candidate paths. As a result, speculative execution
generates far more tokens, which amplifies the computational load of the decode phase. Second,
verification is triggered independently for each candidate path, resulting in frequent,
fine-grained verification tasks. These small and irregular tasks not only underutilize the GPU, but
also repeatedly interrupt the reasoning steps. We provide an in-depth performance
characterization of speculative execution in the following.

\vspace{0.05in}
\noindent {\bf Environment:}
We evaluate test-time scaling with beam search using {\tt Qwen2.5-3B-Instruct} as the generation
model and {\tt Qwen2.5-Math-PRM-7B} as the process reward model. We use beam sizes of 4 and 8.
All experiments are conducted on a single NVIDIA A100 80GB GPU.
We describe the experimental setup in more detail in \autoref{sec:setup}.

\begin{figure}[t]
\centering
\includegraphics[width=3.2in]{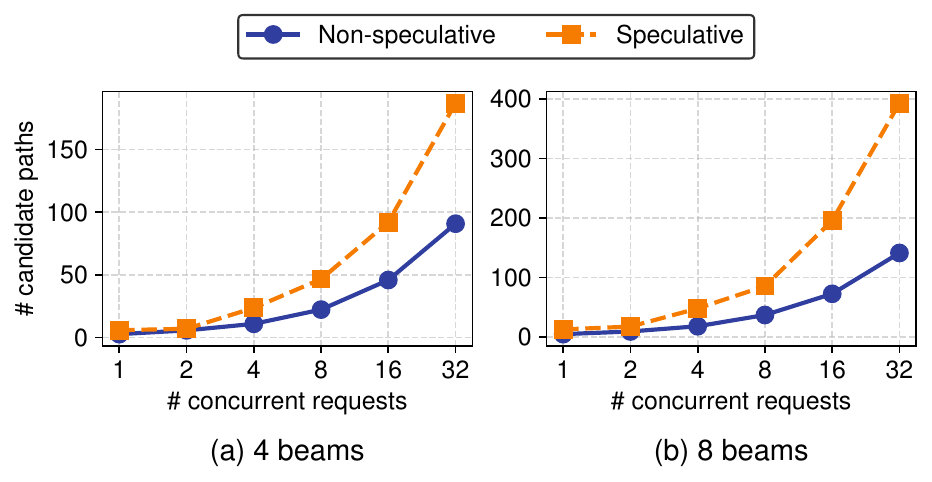}
\vspace{-0.10in}
\caption{Number of candidate paths according to concurrent requests
    on MATH dataset using Qwen2.5-3B-Instruct}
\vspace{-0.10in}
\label{fig:motiv_num_seqs}
\end{figure}

\subsubsection{Increased Decode Computation}
\label{sec:increased_decode}

We first evaluate the additional computational overhead introduced by speculative execution.
\autoref{fig:motiv_num_seqs} presents the number of concurrently explored candidate paths 
(i.e., reasoning sequences)
for both non-speculative and speculative approaches as the number of requests increases.
With a beam size of 4, the non-speculative approach concurrently explores an average of roughly
three active candidates per request, whereas speculative execution increases this to around six.
As the number of requests increases, the gap in the number of simultaneously running candidates
widens significantly. When the beam size is increased to 8, this discrepancy becomes even more
severe, indicating substantially higher computational load under speculative execution.
Speculative execution performs extra computation on candidate paths that do not contribute to
the final result, unlike the non-speculative method, which executes only the selected generation
path. 

\begin{figure}[t]
\centering
\includegraphics[width=3.2in]{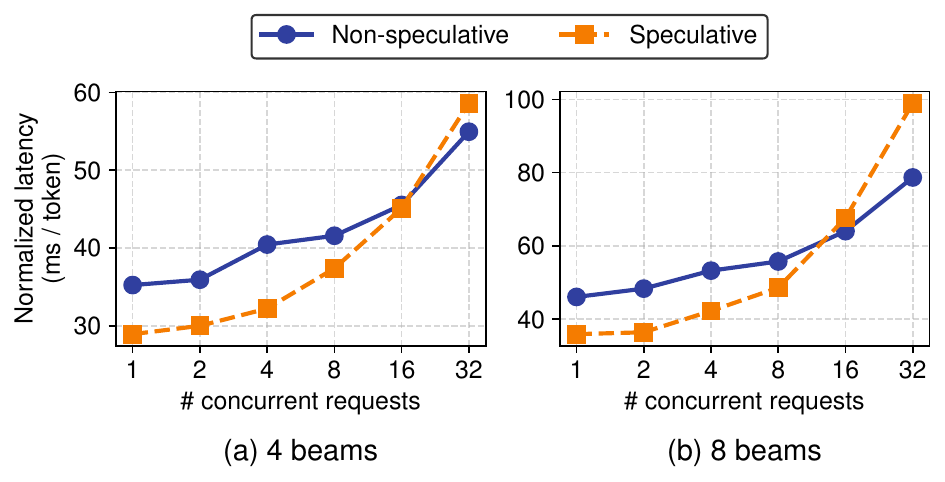}
\vspace{-0.10in}
\caption{Normalized token latency with regard to concurrent requests 
    on MATH dataset using Qwen2.5-3B-Instruct}
\vspace{-0.15in}
\label{fig:motiv_latency}
\end{figure}

Second, we measure average token generation latency to quantify the increased computational load.
\autoref{fig:motiv_latency}
shows the output token latency as the number of requests increases. When the number of requests is low,
speculative execution significantly reduces the output token latency. However, as the number of requests
grows, the latency of the speculative approach increases much more rapidly than that of the baseline.
Eventually, for more than 8 requests in both cases, its latency surpasses the non-speculative approach.

\subsubsection{Increased Verification Steps}
\label{sec:increased_verif}

The speculative approach introduces additional performance overhead in the form of fine-grained verification tasks,
as each candidate path performs verification independently to proceed to the next step. In contrast,
non-speculative execution benefits from bulk verification across candidates within the same step, which
enables efficient batching and high compute utilization. Speculative execution breaks this batching
opportunity, leading to inefficient, fragmented verifications and reduced hardware utilization. 
Note that the verification task is a prefill-only workload, where the PRM verifies the given tokens in
a single forward pass without autoregressive decoding. As verification occurs more frequently, the
generation model is repeatedly preempted, leading to further performance degradation.

\begin{figure}[t]
\centering
\includegraphics[width=3.2in]{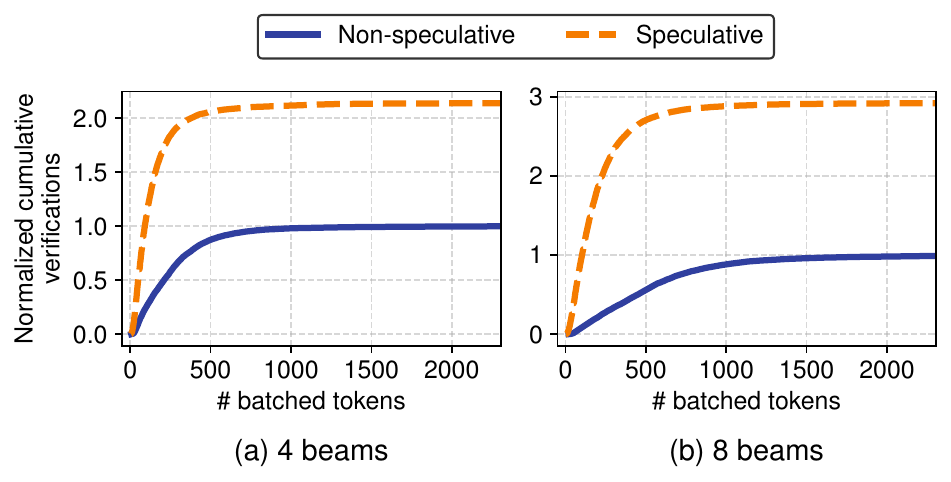}
\vspace{-0.10in}
\caption{Cumulative number of tokens per verification batch on MATH dataset using Qwen2.5-3B-Instruct}
\vspace{-0.2in}
\label{fig:motiv_verif_dist}
\end{figure}

\autoref{fig:motiv_verif_dist} shows the normalized cumulative number of verifications for the non-speculative 
and speculative methods. All values are normalized by the total number of verifications in the non-speculative 
baseline, so the non-speculative curve always converges to 1, while the speculative curve indicates how many 
times more verification calls it performs relative to the baseline. In both beam size settings, the speculative 
method accumulates verifications much more rapidly than the non-speculative baseline within a smaller number of
batched tokens, and its curve eventually saturates at roughly 2--3$\times$ the baseline level.

\section{Towards Efficient Speculation}

\begin{figure}[t]
\centering
\includegraphics[width=3.3in]{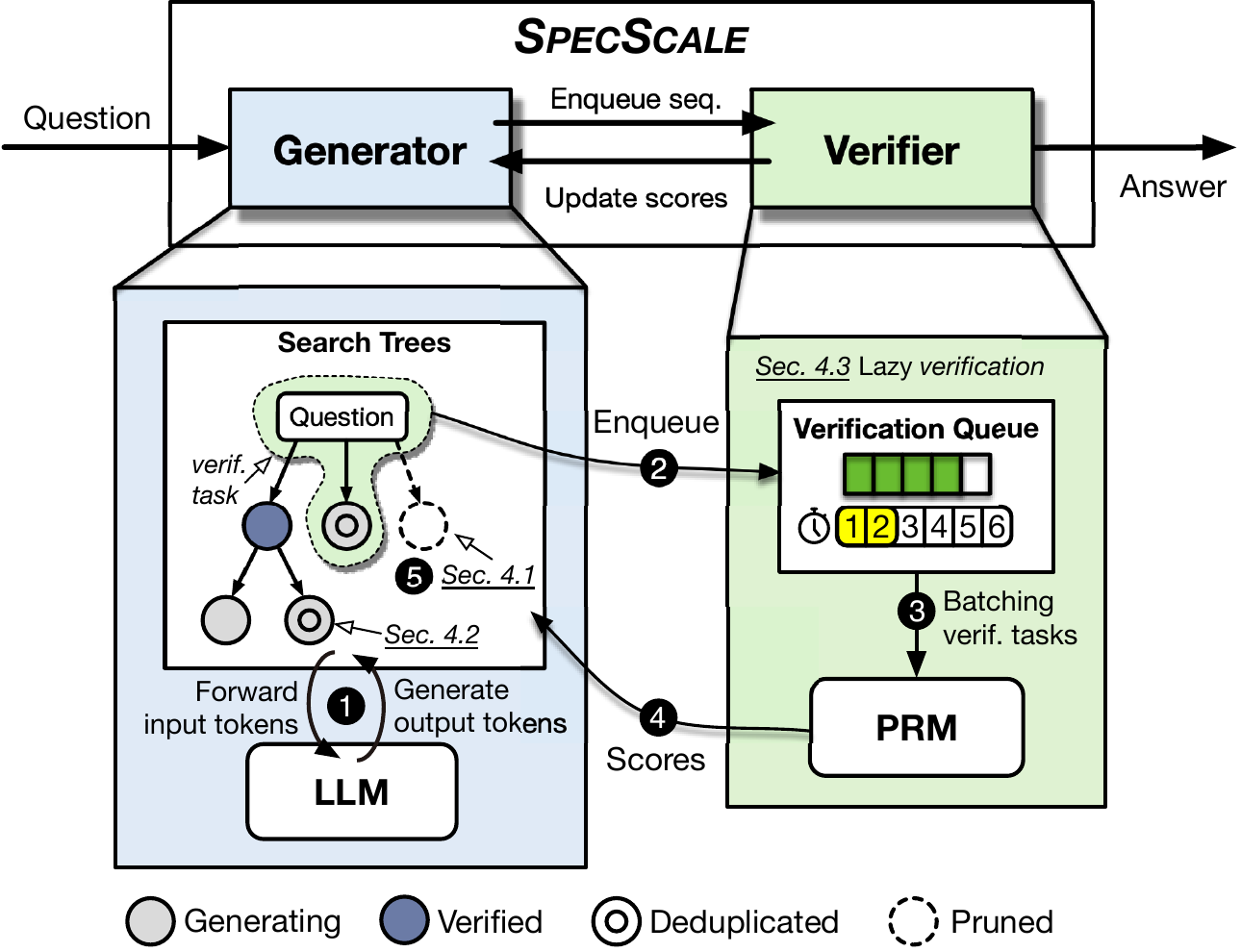}
\vspace{-0.10in}
\caption{Overview of our proposed system: \sys}
\vspace{-0.10in}
\label{fig:overview}
\end{figure}

To address the challenges of speculative execution, we design \sys around three key
techniques: early pruning, computation deduplication, and lazy verification.
\autoref{fig:overview} presents an overview of our proposed system and its workflow. 
\circled{1} When a question arrives, the generator constructs a search tree of
candidate paths using the LLM. At this time, our computation deduplication
technique is applied to reduce the computational overhead. \circled{2} Once a candidate
reaches a verification point (e.g., completes a step), the corresponding node is enqueued
to the verifier as a verification task. \circled{3} The verifier applies lazy verification:
instead of processing each task immediately, it opportunistically batches queued tasks
and runs the PRM on large, GPU-efficient batches. \circled{4} The resulting PRM scores
are sent back to the generator, which updates the search tree and decides which candidates
are allowed to continue speculatively. \circled{5} Using these scores, \sys performs
early pruning, terminating branches that can no longer become part of the final top-$k$
solutions, and focusing computation on the most promising paths.
In the following, we describe these three techniques in turn: early pruning,
computation deduplication, and lazy verification.

\begin{figure*}[t]
\centering
\includegraphics[width=6.3in]{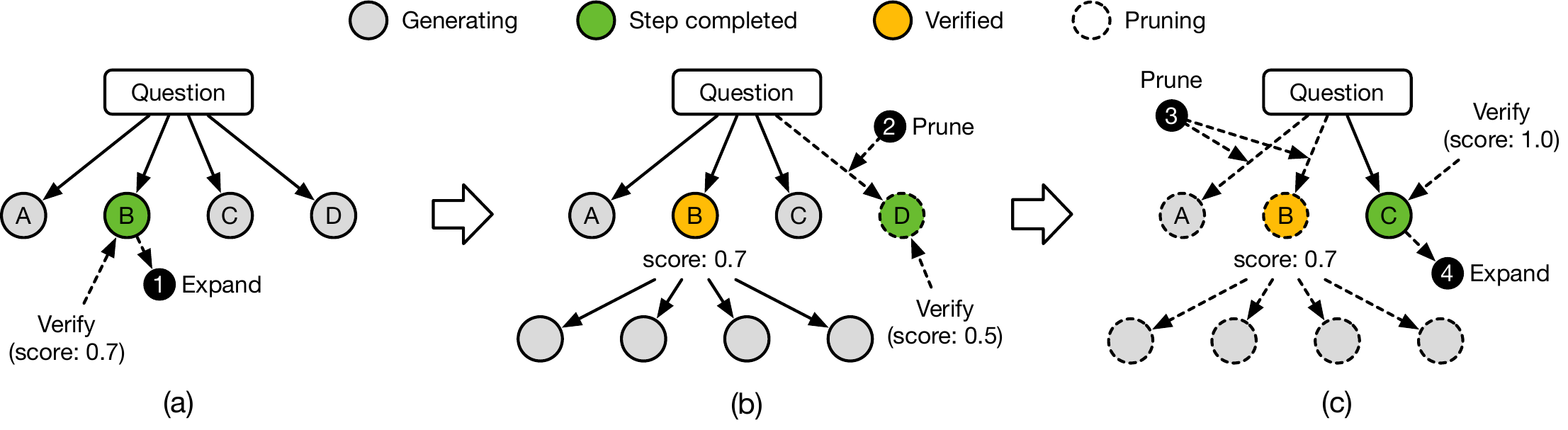}
\vspace{-0.10in}
\caption{Speculative execution with early pruning, where beam size ($N$) and beam width ($M$) are both 4, implying top-$k = 1$}
\vspace{-0.05in}
\label{fig:pruning_approach}
\end{figure*}

\subsection{Early Pruning of Search Tree}

While reasoning with multiple candidate paths, not all candidates are equally promising.
Although the set of top-k candidates can only be finalized after all candidates complete
verification, it becomes possible to determine that some paths can no longer be part of
the final top-$k$ results, even before all verification finishes. Early pruning exploits
this insight by terminating the evaluation of such paths as soon as possible. For example,
once the number of verified candidates exceeds $k$, any candidate whose score is ranked below
the current top-$k$ cannot be part of the final top-$k$ set, regardless of how the remaining
evaluations proceed. Thus, continuing to process such candidates is wasteful.

\autoref{fig:pruning_approach} illustrates the workflow for pruning candidates during
speculative execution. We assume a beam size of 4 and a beam width of 4, which implies
$k = 1$. Suppose that four candidate sequences initialized with a prompt are generating
their first step in parallel. During this process, once a candidate (e.g., {\tt B})
completes the step, the candidate performs verification. \circled{1} After verification,
the candidate {\tt B} expands its candidate path and proceeds to the next step as
speculative execution. Subsequently, when another candidate (e.g., {\tt D}) completes its
step and is verified, if the number of verified candidates exceeds $k$ (i.e., 1)
and this candidate's score is lower than those of the previously verified candidates,
further exploration of its path becomes unnecessary, as at least $k$ other candidates
with higher scores already exist. \circled{2} Thus, we can immediately
prune the lowest-scoring candidate {\tt D} to minimize the additional computation introduced
by speculative execution.

\begin{figure}[t]
\centering
\includegraphics[width=3.2in]{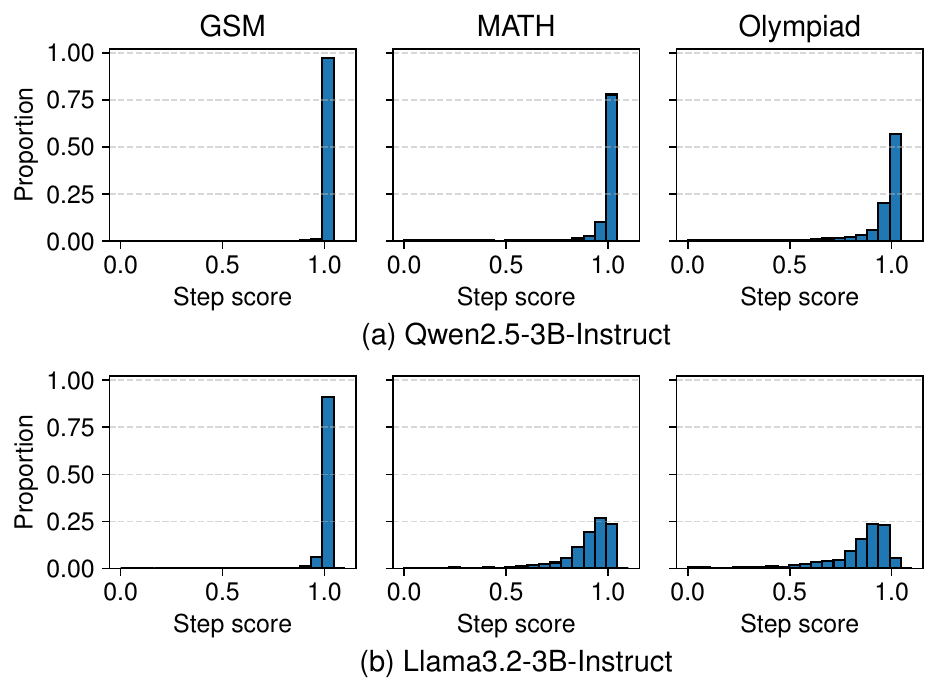}
\vspace{-0.15in}
\caption{Step score distribution}
\vspace{-0.15in}
\label{fig:motiv_step_score}
\end{figure} 

In addition, we leverage the scoring characteristics of the PRM. The PRM produces
floating-point scores in the range $[0, 1]$, where the maximum value 1.0 denotes a perfect score.
When multiple candidates have the same score, we cannot distinguish which one is strictly better~\cite{beeching:2024}.
Consequently, once the number of candidates with a perfect score reaches $k$, we can finalize
them as the selected candidates and prune all remaining ones. For instance, if
a verified candidate {\tt C} receives the perfect score of 1.0,
\circled{3} we prune all remaining candidate paths and \circled{4}
finalize {\tt C} as the selected candidate. 

\autoref{fig:motiv_step_score} presents the distribution of step-wise scores assigned by the
PRMs across the three datasets. In this experiment, we use Qwen2.5-3B-Instruct and 
Llama3.2-3B-Instruct as generative models, together with Qwen2.5-Math-PRM-7B and
Llama3.1-8B-PRM as their corresponding PRMs, respectively. Even with the small 
model Qwen2.5-3B-Instruct for the Olympiad dataset, a large fraction of steps receive
the maximum score of 1.0, indicating that we can significantly reduce unnecessary speculation
with early pruning.

\subsection{Computation Deduplication}
\label{sec:cd}
One key inefficiency in exploring multiple candidates is redundant decoding across 
candidate paths that share the same token prefix. In existing LLM serving systems, multiple
beams (i.e., candidate paths) are instantiated as independent sequences, all initialized
with the same prompt and subsequently expanded in parallel.

\begin{figure}[t]
\centering
\includegraphics[width=3.3in]{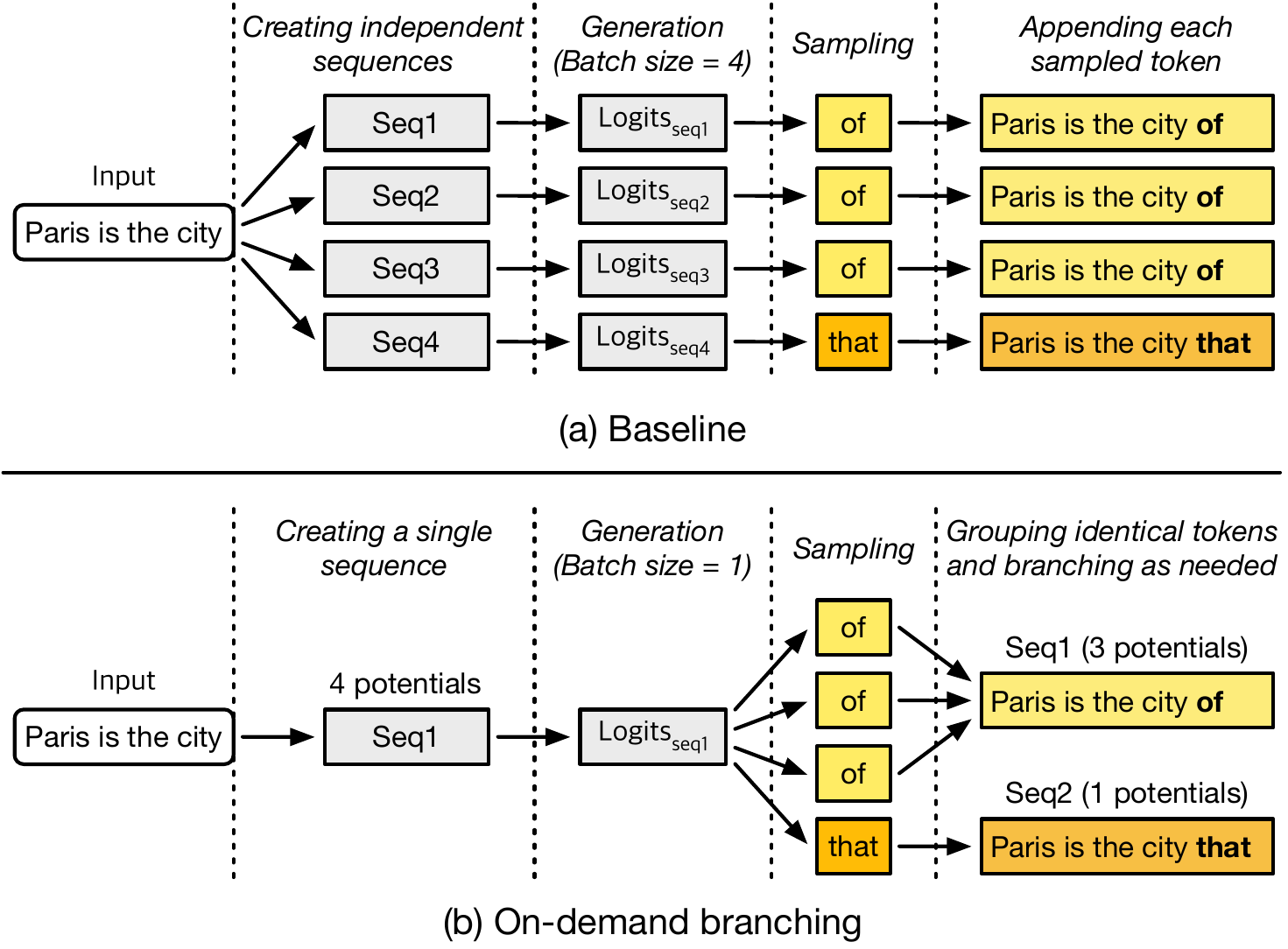}
\vspace{-0.15in}
\caption{Baseline vs. computation deduplication approach}
\vspace{-0.2in}
\label{fig:dedup_approach}
\end{figure}

\autoref{fig:dedup_approach}a
shows an example with a beam size of 4, where the model executes four separate forward 
passes, even though they all evaluate the logits for the same input prefix, {\it ``Paris is
the city''}. As a result, each decoding iteration recomputes the same logits separately
for every candidate. Moreover, in this example, three paths sample the same next token,
{\it ``of''}, incurring redundant computation again.

\begin{figure}[t]
\centering
\includegraphics[width=3.2in]{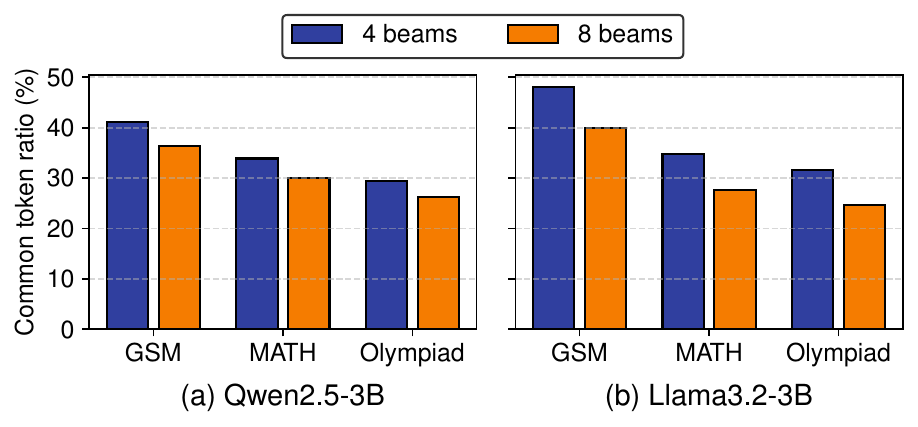}
\vspace{-0.10in}
\caption{Common token (\%) across candidates of each step}
\vspace{-0.20in}
\label{fig:motiv_common_token}
\end{figure}

To quantify the degree of redundancy, we measure the common prefix token ratio.
In this experiment, we use a sampling temperature of 0.8, following the setting
in ~\cite{beeching:2024}. Across the three datasets, \autoref{fig:motiv_common_token}
shows the fraction of tokens that can be
shared among completed candidates at each step. We observe that 25--50\% of generated tokens
are common across candidates. The ratio is higher on easier benchmarks, where the model’s
predictions are more confident, indicating that similar reasoning processes take place
across candidates. Overall, these results reveal substantial opportunities for sharing
computation, even under relatively diverse sampling regimes.

To eliminate redundant computation, we propose a deduplication method that shares output
logits among the candidates with identical input sequences. 
\autoref{fig:dedup_approach}b illustrates how the proposed method
effectively eliminates redundant computation via {\it on-demand branching}.
Instead of instantiating four independent sequences, we create a single sequence whose
{\tt potential} size is initialized to 4, indicating that it implicitly represents four
candidates. When this sequence is scheduled, we run a single decoding step over the
current prefix to obtain the output logits. We then sample as many tokens as the potential
size (4 in this example) from the same logits. Note that each token is sampled as an
independent trial (i.e., four times) from the same distribution. 

After sampling, we identify the opportunity to group the sampled tokens.
For instance, if three samples produce {\it ``of''} and one produces {\it ``that'' }, we form two
groups: \{{\it ``of'', ``of'', ``of'' }\} and \{{\it ``that'' }\}. Once more than one group is formed,
we create one child sequence per group, extending each with its corresponding token and assigning
its potential size to the group’s cardinality (3 and 1 in the example). 

This construction yields a tree of explicit sequences that is probabilistically equivalent
to running four independent samples, but it executes exactly one model forward pass for each
distinct token prefix. The amount of decode computation is therefore proportional to the number of unique
prefixes explored, rather than to the total number of logical candidates.
This differs from prefix-based KV sharing in existing serving systems,
which reduces redundant KV computation over shared prefixes~\cite{zheng:2024}
but still requires one forward pass per candidate. {\tt CD} instead performs a single forward pass
for identical candidates and samples all corresponding candidates from the same logits.

Meanwhile, when sampling next tokens, each sequence can have a different potential size,
indicating that it needs a different number of samples from its logits. As a result, we
cannot sample all sequences in a single batch, because each one would require a different
output shape. To preserve batching efficiency, we group sequences by potential size and
perform batched sampling separately for each group.

\subsection{Lazy Verification}
\label{sec:lazy_verif}

To mitigate the verification overhead caused by speculative execution, we introduce a deferred
batching strategy that queues verification requests and processes them in large, GPU-efficient
batches. 
The basic idea is simple. Instead of invoking the verification immediately whenever a candidate
completes a step, we enqueue its verification request in a buffer. Periodically, we flush the
buffer and run verification on a batch containing many candidates at once. This improves throughput
because verification is a prefill-only workload whose performance scales well with batch size.

However, deferring verification introduces an inherent latency-throughput trade-off: longer
deferrals yield larger, more efficient batches but reduce the benefit of speculative execution.
Our lazy verification identifies a balanced point that reconciles high verification
throughput with low idle time for efficient speculative execution. We use three policies that
jointly control the verification time.

\begin{figure}[t]
\centering
\includegraphics[width=3.2in]{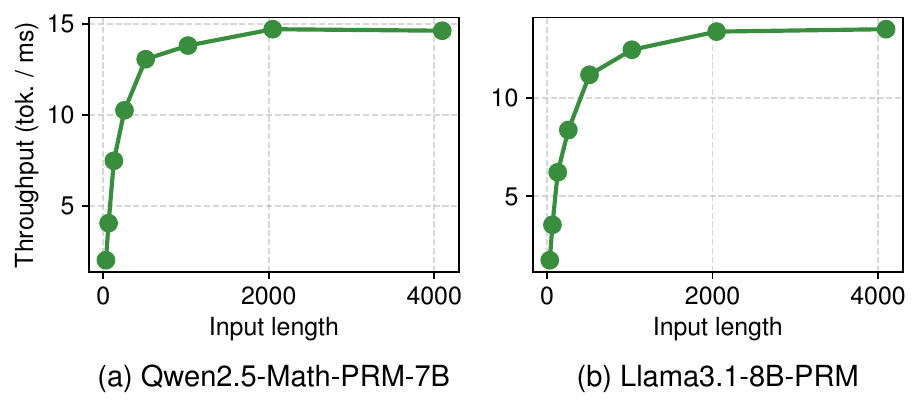}
\vspace{-0.10in}
\caption{Throughput of verification in a microbenchmark}
\vspace{-0.2in}
\label{fig:analysis_verif_latency}
\end{figure} 

First, we flush the queue when the total number of tokens across all enqueued requests exceeds
a pre-defined threshold. We determine this threshold via offline profiling. Specifically, we
measure the verification token throughput of two PRM models, Qwen2.5-Math-PRM-7B and Llama3.1-8B-PRM,
as we vary the number of input tokens.
\autoref{fig:analysis_verif_latency} shows that token throughput saturates around 2,048 input
tokens for both models: beyond this point, additional tokens yield diminishing throughput gains. 
For other models or hardware configurations, the same profiling step can
be run once before deployment to select an appropriate threshold.

Second, when all candidates within the same sibling group (i.e., descendants of the same parent
node in the search tree) complete their current step, we immediately verify them together,
along with any deferred requests currently in the queue. Further delaying these candidates
would introduce extra latency compared to the non-speculative baseline without providing
additional batching benefit. 

Finally, under low request load, the verification queue may take a long time to accumulate
enough tokens to reach the threshold, which again increases the waiting time for the early
completed paths. To prevent excessive deferral in this regime, we track, for each sample,
how many times its verification steps have been deferred. Once this per-sample deferral
count exceeds a small limit, we flush the queue and immediately verify all pending candidates.
In our deployment, we empirically set the maximum deferral count to six, which provides
stable latency under light load.

These policies allow \sys to opportunistically batch verification workloads,
improving GPU utilization and reducing the number of fragmented verification calls, while
keeping verification latency within reasonable bounds.

\section{Evaluation}
\subsection{Experimental Setup}
\label{sec:setup}

\noindent {\bf Environment and models:}
To evaluate \sys and baselines on a common serving substrate, we implement a lightweight framework
that incorporates state-of-the-art LLM serving optimizations, including
continuous batching~\cite{yu:2022}, paged attention~\cite{kwon:2023},
prefix KV caching and sharing~\cite{zheng:2024}, and fused attention kernels based on FlashInfer~\cite{ye:2025}.
We run all experiments using this framework on a server equipped with Intel Xeon Gold 6326 processors
and a single NVIDIA A100 80GB GPU. We use PyTorch v2.9~\cite{pytorch} and CUDA 12.8~\cite{cuda}.

We pair instruction-tuned LLMs as generators with process
reward models (PRMs) as verifiers. Specifically, we use two models from the Qwen2.5
family (3B and 7B) and two from the Llama family (3B and 8B) as generators.
For verification, we adopt the corresponding PRMs trained for mathematical reasoning
tasks: Qwen2.5-Math-PRM-7B for the Qwen models and Llama3.1-8B-PRM for the Llama models.
This pairing ensures architectural consistency between each generator–verifier pair.

\vspace{0.04in}
\noindent{\bf Benchmarks:}
For performance evaluation, we use three math reasoning datasets:
GSM8K~\cite{gsm}, MATH-500~\cite{math}, and OlympiadBench~\cite{olympiad}.
For GSM8K, we randomly sample 1,000 problems and use all problems from MATH-500 and OlympiadBench.
We set the maximum number of concurrently running requests to 32, because this is where the non-speculative baseline
starts to surpass the naive speculation (\autoref{sec:increased_decode}).

We follow the hyperparameter configuration of Snell et al.~\cite{snell:2025}
for our beam search-based test-time scaling, including a cap of 40 beam expansion steps and
a fixed beam width of 4. Since their configuration does not specify the sampling
temperature, we set it to 0.8, following Beeching et al.~\cite{beeching:2024}.

\begin{figure*}[t]
\centering
\includegraphics[width=6.9in]{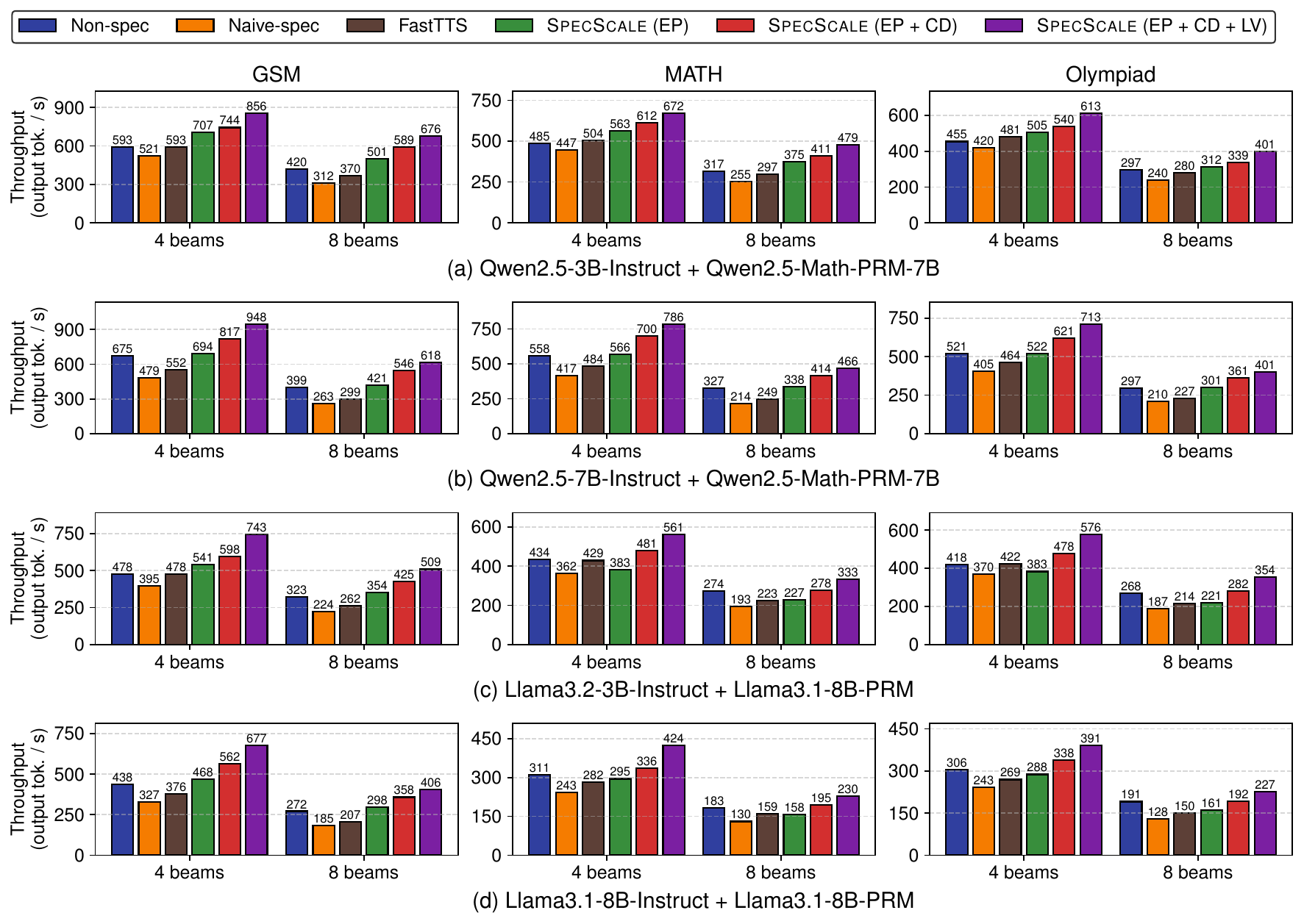}
\vspace{-0.10in}
\caption{Token throughput performance for GSM, MATH, and Olympiad datasets on Qwen2.5 and Llama3 models}
\vspace{-0.1in}
\label{fig:eval_throughput}
\end{figure*}

\vspace{0.04in}
\noindent{\bf Comparisons:}
We evaluate the effectiveness of our techniques against three baselines,
a conventional beam search method without speculation ({\tt Non-spec}),
a naive speculation ({\tt Naive-spec}), and {\tt {FastTTS}}~\cite{chen:2025}, a recently proposed approach.
{\tt Naive-spec} expands multiple candidate paths through speculative execution
and performs verification for each completed step, resulting in frequent verification.
{\tt FastTTS} mitigates this overhead through two techniques, Speculative Candidate
Selection and Lookahead Verification, both of which we reimplement in our framework
because the publicly released code supports only a batch size of one
~\cite{fasttts:artifact}
and thus cannot accommodate our target serving setting.

Both techniques differ from our design. Lookahead Verification batches completed beams
only within a single request, whereas our lazy verification batches them across concurrent
requests once a flush is triggered, yielding larger verification batches and higher GPU
utilization. Speculative Candidate Selection reduces unnecessary expansions, but still
suffers from redundant computation. 

We omit the other two techniques of {\tt FastTTS}, Preemptible Scheduling and Dynamic
Prefix-Aware Scheduling. They target severely resource-constrained deployments in which 
the beams of even a single request cannot all be executed concurrently. Our study instead
focuses on a server-class environment handling multiple concurrent requests. 

To validate our reimplementation, we compared its throughput with the released FastTTS implementation at a batch size of one, using Qwen2.5-7B-Instruct on MATH-500. The two implementations achieved similar throughput at beam sizes of 4, 8, 16, and 32; detailed results are omitted due to space constraints.

\subsection{Throughput Performance}
We first evaluate the maximum output token throughput across three datasets and
four model configurations.
To assess the contribution of each component, we incrementally
enable our three techniques, early pruning ({\tt EP}), computation deduplication
({\tt CD}), and lazy verification ({\tt LV}), on top of the naive speculation method.
For this evaluation, we issue all requests
immediately and measure the token throughput over all problems in the datasets.

\autoref{fig:eval_throughput} presents token throughput for beam sizes 4 and 8. In most cases,
the naive speculation approach results in lower token throughput than the non-speculative case.
This is because naive speculation increases the amount of decode computation as well as verification
calls, as discussed in \autoref{sec:chall_spec}. In a serving environment, the performance benefits
of naive speculative execution exhibit diminishing returns due to frequent verification calls.
FastTTS alleviates the frequent verification overhead of naive speculation by batching verification calls,
leading to improved token throughput. Although FastTTS reduces the number
of active candidates by dynamically adjusting the branching factor,
its benefits are still insufficient for large models in serving environments
where multiple requests are processed concurrently.

On the other hand, \sys ({\tt EP}) achieves better performance compared to the naive-speculative approach by effectively
mitigating unnecessary speculative execution, but it shows performance similar to the non-speculative
case.
On the MATH and Olympiad datasets with 8 beams, \sys ({\tt EP}) achieves 1.18$\times$ and
1.05$\times$ higher throughput, respectively, than the non-speculative baseline on the Qwen2.5-3B-Instruct
model. For the Llama-3.2-3B-Instruct model, although early pruning improves throughput over the
naive speculative approach, it still does not reach the non-speculative baseline.
This discrepancy mainly comes from how frequently each model produces perfect scores during verification.
When a candidate with a perfect score is generated, we can immediately advance to the next step, 
thereby bypassing additional verification and reducing overall verification overhead.
However, as shown in \autoref{fig:motiv_step_score}, the Llama models generate fewer perfect scores than Qwen2.5.
Consequently, the verification overhead is not sufficiently reduced, and the throughput benefit of speculative execution is limited for Llama models.

\sys ({\tt EP+CD}), which additionally incorporates computation deduplication, further improves performance.
On the MATH dataset with 8 beams, \sys ({\tt EP+CD}) further improves throughput by 1.10$\times$
and 1.22$\times$ over \sys ({\tt EP}) on the Qwen2.5-3B-Instruct and Qwen2.5-7B-Instruct models, respectively.
The additional performance gain of {\tt EP+CD} is less significant for smaller models than for larger ones, because small models do
not fully utilize the available compute resources even with speculative execution, making the
benefit of reducing redundant computation less significant. Finally, \sys ({\tt EP+CD+LV}), which
includes all three techniques, yields the highest token throughput among all configurations by
further reducing verification overhead. 

Overall, \sys consistently delivers higher throughput than both the non-speculative baseline and FastTTS.
On the MATH dataset, with 8 beams, \sys achieves 1.51$\times$ and 1.43$\times$ higher throughput
than the baseline for Qwen2.5-3B-Instruct and Qwen2.5-7B-Instruct, respectively.

\begin{figure*}[t]
\centering
\includegraphics[width=6.8in]{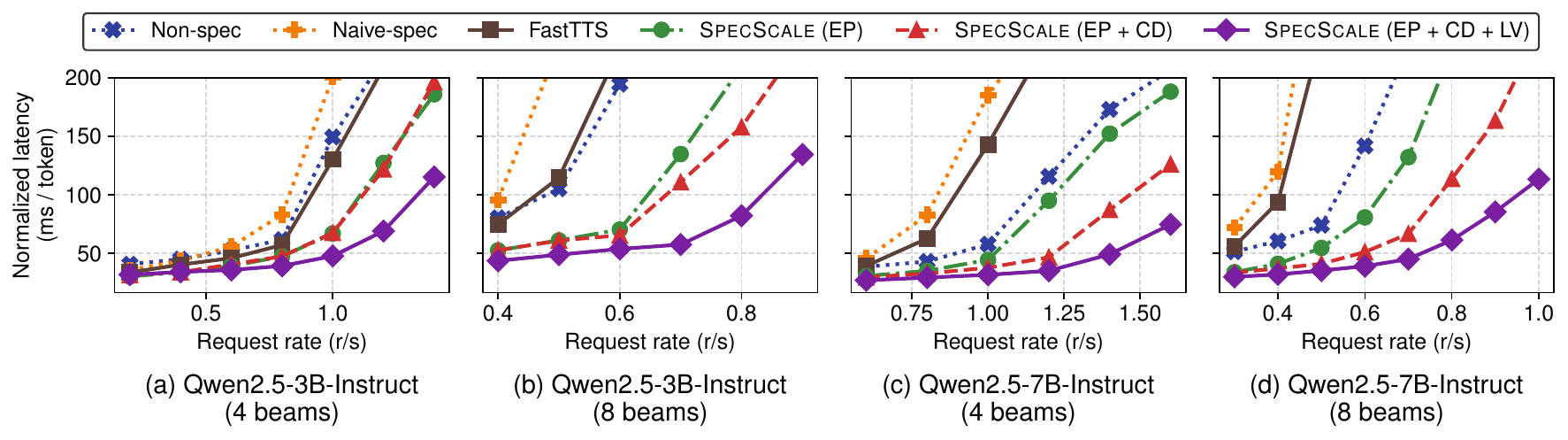}
\vspace{-0.1in}
\caption{Serving performance for MATH on Qwen2.5-3B-Instruct and Qwen2.5-7B-Instruct}
\vspace{-0.1in}
\label{fig:eval_serving}
\end{figure*}

\subsection{Serving Performance}
In this section, we evaluate the serving performance of models utilizing test-time scaling methods.
We measure the end-to-end token latency of Qwen2.5-3B-Instruct and Qwen2.5-7B-Instruct
by increasing the request rate in our serving framework. We present normalized output token latency,
defined as the end-to-end latency divided by the number of generated output tokens~\cite{yu:2022}.
We use Poisson distributions for generating a realistic request arrival pattern.

\autoref{fig:eval_serving} exhibits the normalized output token latency (y-axis) against the
request rate (x-axis) for beam sizes 4 and 8 on the MATH dataset.
The naive speculation approach shows comparable latency to, or even slightly lower than,
the non-speculative case at low request rates. This is because, despite effectively reducing
waiting time, it triggers frequent small verification batches, which introduce verification
overhead and thereby offset the latency gains. At higher request loads, its performance further degrades
as the increased number of candidates amplifies the overall computational overhead,
which becomes more significant with larger beam sizes or larger models.

FastTTS exhibits lower latency than both non-speculative and naive speculative approaches on Qwen2.5-3B-Instruct
by reducing the verification overhead through Lookahead Verification.
However, as the request rate increases, the performance gap narrows. With larger beam sizes or larger models,
FastTTS leads to higher latency than the non-speculative approach, as the computational overhead of speculative execution dominates even with Speculative Candidate Selection.

In contrast, \sys ({\tt EP}) effectively mitigates unnecessary speculative executions
and improves the normalized latency. \sys ({\tt EP+CD}) further enhances the normalized latency
by eliminating redundant computation. The performance benefit of computation
deduplication is not observed for Qwen2.5-3B-Instruct (4 beams), as the smaller beam size does not
fully utilize the computational resources. On the other hand, larger beam sizes or larger models show the 
effectiveness of deduplication. Finally, \sys ({\tt EP+CD+LV}) achieves the lowest
normalized latency among all configurations by substantially reducing verification overhead.

For Qwen2.5-7B-Instruct with 4 beams, our {\tt EP+CD+LV} achieves an average normalized
latency of about 75 ms at a load of roughly 1.6 requests per second. At a similar latency
level, FastTTS can serve only around 0.8 requests per second.
For a given latency budget, \sys can serve more requests (i.e., higher goodput) than
both the naive speculation and FastTTS. This allows \sys to explore more candidates
(i.e., more beams) within the same budget, potentially leading to higher reasoning accuracy.

\begin{figure*}[t]
\centering
\includegraphics[width=7.in]{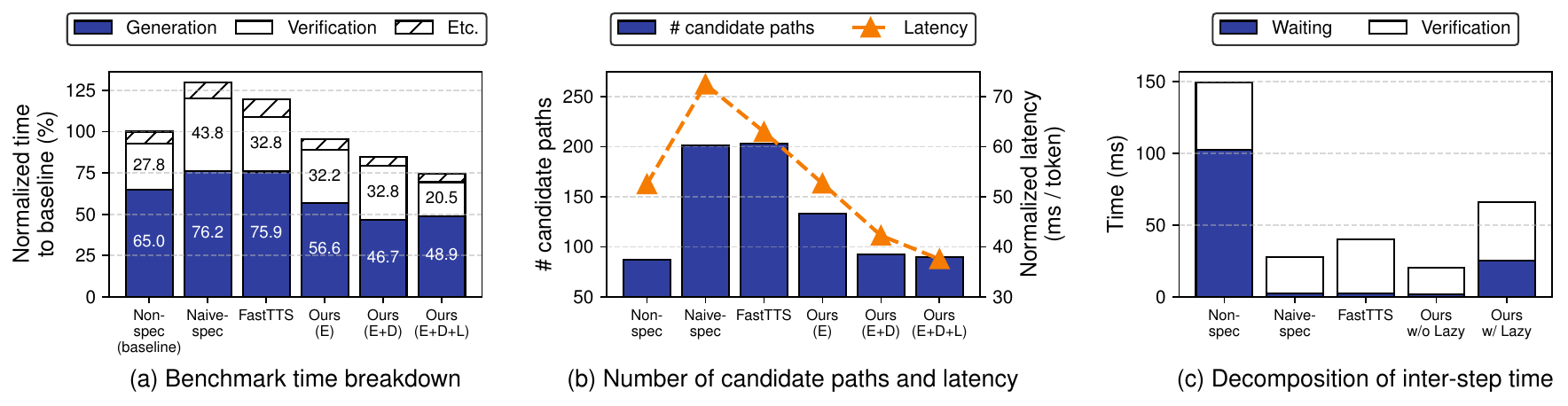}
\vspace{-0.2in}
\caption{Quantitative analysis of \sys and baselines for MATH on Qwen2.5-7B-Instruct with beam size 4
({\tt \textbf{E}}: Early Pruning, {\tt \textbf{D}}: Computation Deduplication, and {\tt \textbf{L}}: Lazy Verification)}
\label{fig:eval_overall}
\end{figure*}

\subsection{Performance Analysis}

To analyze the performance improvements, we quantify the contributions of our three
techniques by comparing them against the baselines. For this analysis, we focus on
Qwen2.5-7B-Instruct model with a beam size of 4 on the MATH dataset, corresponding to the
throughput results in \autoref{fig:eval_throughput}. 

\autoref{fig:eval_overall}a shows where execution time is spent
by breaking it into three components: generation, verification, and the remaining time
(labeled as {\tt Etc.}, including scheduling and tokenization). In the non-speculative case, generation
takes approximately 2.3$\times$ longer than verification, and {\tt Etc.} accounts for about
7\% of the total runtime.  Within {\tt Etc.}, tokenization alone contributes roughly 40\% of
the time, because the PRM requires a specific input template, forcing us to reformat the
generated output tokens at every verification step. For naive speculation, both generation
and verification take substantially longer than in the baseline, leading to even higher
overall runtime. In FastTTS, the verification time is reduced, but the generation time remains significant,
as the dynamic branching is ineffective at small beam sizes.
In particular, at a beam size of 4, it is not applied because candidates share the same previous scores.

\begin{table*}[t]
\footnotesize
\centering
\begin{subtable}[t]{0.54\linewidth}
\centering
\begin{tabular}{l|c|c|c|c}
\toprule
\multirow{2}{*}{Dataset}   & \multirow{2}{*}{Non-spec} & \multicolumn{3}{c}{\sys} \\
\cmidrule{3-5}
     &  & {\tt EP} & {\tt EP+CD} & {\tt EP+CD+LV} \\
\midrule
GSM8K         & 92.99 & 93.90 ($+0.91$)   & 93.79 ($+0.80$) & 93.49 ($+0.50$) \\
MATH-500      & 74.03 & 73.96 ($-0.07$)  & 73.97 ($-0.06$) & 75.43 ($+1.40$) \\
OlympiadBench & 27.18 & 27.51 ($+0.33$)  & 28.64 ($+1.46$) & 27.94 ($+0.76$) \\
\bottomrule
\end{tabular}
\vspace{0.05in}
\caption{4 beams}
\label{tab:eval_accuracy_beam4}
\end{subtable}
\hfill
\begin{subtable}[t]{0.42\linewidth}
\centering
\begin{tabular}{c|c|c|c}
\toprule
\multirow{2}{*}{Non-spec} & \multicolumn{3}{c}{\sys} \\
\cmidrule{2-4}
      & {\tt EP} & {\tt EP+CD} & {\tt EP+CD+LV} \\
\midrule
94.70 & 94.20 ($-0.50$) & 93.80 ($-0.90$) & 94.09 ($-0.61$) \\
76.24 & 79.21 ($+2.97$) & 79.16 ($+2.92$) & 76.82 ($+0.58$) \\
30.40 & 29.39 ($-1.01$) & 30.70 ($+0.30$) & 31.48 ($+1.08$) \\
\bottomrule
\end{tabular}
\vspace{0.05in}
\caption{8 beams}
\label{tab:eval_accuracy_beam8}
\end{subtable}
\caption{Answer accuracy (\%) of Qwen2.5-3B-Instruct with a PRM verifier. The numbers in parentheses denote the accuracy difference in percentage points compared to the baseline ({\tt Non-spec}).}
\vspace{-0.20in}
\label{tab:eval_accuracy}
\end{table*}

On the other hand, \sys ({\tt EP}) reduces both generation and verification time.
\autoref{fig:eval_overall}b presents the number of candidate paths (i.e., reasoning sequences)
and the token latency for the baselines and our three
design options. These results are extracted from the same experiment as \autoref{fig:eval_overall}a.
Our early pruning ({\tt EP}) effectively reduces the number of candidate paths by pruning unnecessary
exploration paths early. As a result, the token latency is also reduced compared to naive
speculation. \sys ({\tt EP+CD}) further reduces the generation time by deduplicating computation
across candidates that share identical input sequences. As shown in
\autoref{fig:eval_overall}b, this decreases the number of concurrent
reasoning paths, which in turn lowers the token latency.

While \sys ({\tt EP+CD+LV}) slightly increases the waiting time because verification
is deferred and performed in batches, it reduces the overall verification time by making
the verification process more efficient. \autoref{fig:eval_overall}c
decomposes the inter-step time into waiting time and verification time. Without lazy
verification, the speculative approaches eliminate most of the waiting time, but they suffer from
frequent, small verification calls. With lazy verification, \sys ({\tt EP+CD+LV}) introduces a
modest amount of waiting by deferring verification into batches, yet substantially reduces overall
verification time by making PRM execution more efficient.
Overall, \sys ({\tt EP+CD+LV}) achieves the lowest token latency by improving verification efficiency.

\subsection{Impact on Answer Quality}

We evaluate whether our three techniques affect answer quality.
Early pruning uses two strategies:
1) stopping the speculative expansion of candidates once their
scores indicate that they can no longer enter the final top-$k$, and
2) proceeding to the next step once $k$ candidates achieve
a perfect score.

The first strategy exploits the fact that standard beam search
retains only the top-$k$ candidates at each step.
A candidate whose score falls below the current
$k$-th highest score cannot survive selection.
Thus, stopping the speculative expansion of such candidates
does not affect the final candidate selection.
In contrast, the second strategy may affect candidate selection
because it proceeds as soon as $k$ candidates achieve a perfect
score, even if some remaining candidates would also achieve
the same score and require tie-breaking.

\autoref{tab:eval_accuracy} reports the answer accuracy of
\sys and the non-speculative baseline using
Qwen2.5-3B-Instruct at beam sizes 4 and 8.
These results are extracted from the experiments in
\autoref{fig:eval_throughput}(a).
Across both beam sizes, the accuracy difference of {\tt EP} ranges from
$-1.01$ to $+2.97$ percentage points.
Further incorporating {\tt CD} and {\tt LV} does not materially affect accuracy.
These results indicate that \sys maintains answer accuracy
comparable to the non-speculative baseline.

\subsection{Analysis of Lazy Verification Technique}

To assess our choice of the token threshold for triggering lazy verification,
we conduct a sensitivity study using Qwen2.5-3B-Instruct on
MATH-500, varying the threshold from 256 to 4,096 tokens.
We keep {\tt EP} and {\tt CD} enabled and all other settings
fixed throughout the study.

\begin{figure}[t]
\centering
\includegraphics[width=3.3in]{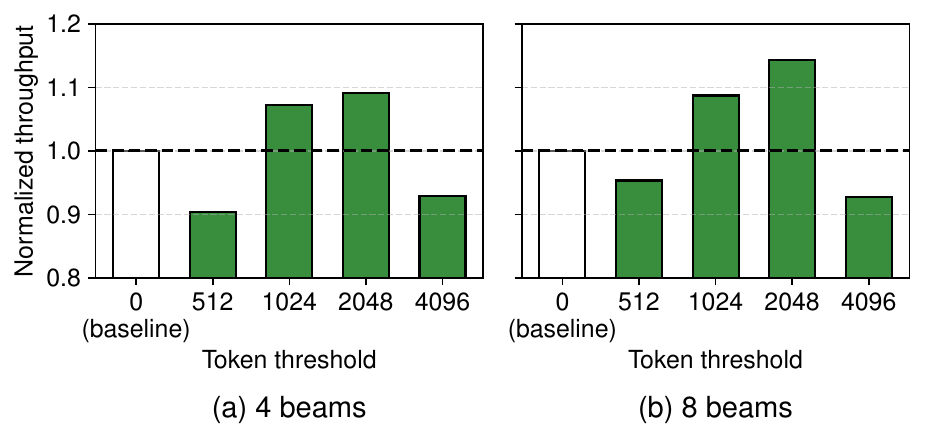}
\vspace{-0.10in}
\caption{Normalized output token throughput with varying batching token thresholds for lazy verification}
\vspace{-0.10in}
\label{fig:eval_lv_sensitivity}
\end{figure}

\autoref{fig:eval_lv_sensitivity} shows output token throughput with
beam sizes of 4 and 8, normalized to the throughput of \sys
with lazy verification disabled ({\tt EP+CD}) for each beam size.
For both beam sizes, throughput improves as the threshold
increases from 512 to 2,048 tokens.
At a threshold of 2,048 tokens, \sys achieves the highest
throughput for both beam sizes, reaching 1.09$\times$ and
1.14$\times$ the baseline throughput for beam sizes 4 and 8,
respectively.
However, further increasing the threshold to 4,096 tokens
reduces throughput below the baseline for both beam sizes.
This trend is consistent with the trade-off discussed in
\autoref{sec:lazy_verif}: larger batches improve verification
efficiency, but excessive deferral reduces the benefit of
speculative execution.

\begin{figure}[t]
\centering
\includegraphics[width=3.3in]{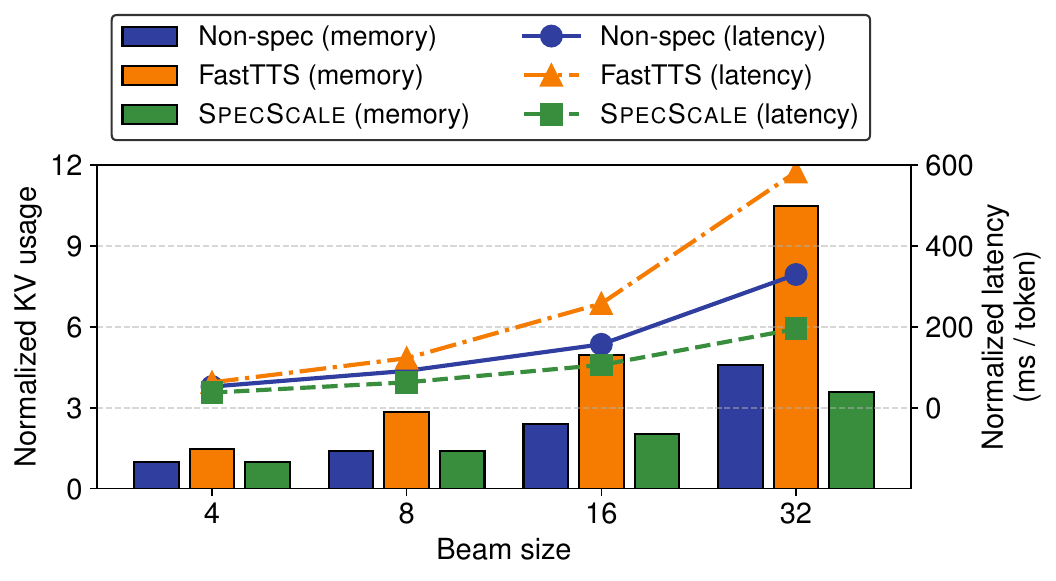}
\vspace{-0.14in}
\caption{Normalized KV memory usage (left y-axis) and token latency (right y-axis) as the beam size increases}
\vspace{-0.14in}
\label{fig:eval_beam_size}
\end{figure}

\subsection{Scalability}

We evaluate how \sys scales with the beam size and concurrency for MATH on Qwen2.5-7B-Instruct. 

\vspace{0.04in}
\noindent{\bf Beam size:}
\autoref{fig:eval_beam_size} shows the normalized output token latency and KV memory
usage as the beam size increases from 4 to 32. For the non-speculative case, the memory usage
for KVs grows approximately in proportion to the beam size because memory allocated to all candidate
paths is not reclaimed until their verification completes.
The latency also increases with the beam size, as more candidate paths are explored and verified in parallel.
For FastTTS, both memory usage and latency increase more sharply due to the speculatively expanded paths,
even though Speculative Candidate Selection becomes effective from a beam size of 8.

On the other hand, \sys reduces active KV usage by early pruning unpromising candidates
and keeps latency lower even as the beam size increases. At a beam size of 32, Non-spec shows
approximately 330ms, whereas \sys maintains a lower latency of about 195ms while
simultaneously reducing KV memory usage by roughly 22\%. For a fixed latency budget, \sys
can afford to use a larger beam size and thus explore more candidates than the baselines.

\begin{figure}[t]
\centering
\includegraphics[width=3.2in]{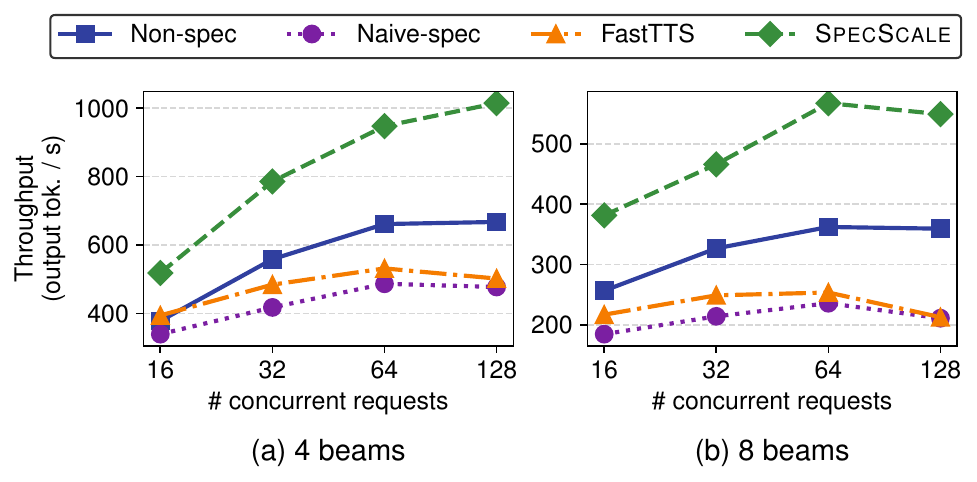}
\vspace{-0.10in}
\caption{Token throughput according to the number of concurrent requests}
\vspace{-0.20in}
\label{fig:eval_batch_size}
\end{figure}

\noindent{\bf Concurrency:} 
\autoref{fig:eval_batch_size} shows the output token throughput when varying the number of concurrent
requests from 16 to 128 with beam sizes of 4 and 8. For a beam size of 4,
both the non-speculative approach and FastTTS saturate around
64 concurrent requests, whereas \sys continues to increase throughput beyond this point.
From 16 to 128 concurrent requests, \sys exhibits higher throughput than the other three approaches for both beam sizes 4 and 8.

\section{Related Work}

Prior studies have addressed the performance overhead of speculative execution in test-time
scaling~\cite{chen:2025, cemri:2025, sixu:2025}.
Recently, FastTTS~\cite{chen:2025} introduced two techniques, Speculative Candidate Selection and Lookahead Verification,
to reduce the overhead of speculative execution and verification, respectively. Speculative Candidate Selection proactively
limits the expansion of candidates predicted to be less promising, whereas our early pruning operates on scores observed
after verification and deterministically eliminates candidates once they can no longer enter the final top-$k$.
Lookahead Verification defers candidate verification, similar to \sys. However, it flushes deferred candidates
only when all candidates complete their current step, which limits opportunities for early pruning.

Concurrent with our work, SPEX~\cite{zhong:2026} addresses the performance bottleneck of synchronized
verification in tree-based test-time compute scaling.
It speculatively expands completed candidates while waiting for stragglers, distributes speculative budget across concurrent requests,
and employs adaptive early termination based on the confidence margin of the leading candidate.
SPEX relies on a heuristic confidence-margin threshold for early termination,
whereas {\tt EP} deterministically discards candidates once $k$ verified candidates outrank them.
In addition, SPEX does not explicitly optimize verification batch
efficiency or redundant forward computation.
{\tt LV} batches fragmented PRM verification across concurrent requests,
while {\tt CD} eliminates redundant forward passes by coalescing candidates that are identical at runtime.

SPECS~\cite{cemri:2025} and ORCHES~\cite{sixu:2025} introduce an additional lightweight model.
SPECS drafts reasoning steps with a small model along high-confidence paths and switches to
the target model depending on problem difficulty. ORCHES proceeds with the scores of a
lightweight verifier while the target PRM verifies in parallel on a GPU-PIM system. This 
requires additional hardware to support concurrent generation and verification and still incurs
misprediction penalties.

Prefix KV caching~\cite{gao:2024, jeong:2025, yao:2025} and shared-prefix attention kernels~\cite{juravsky:2024, yi:2026}
reuse the KV cache of a common prefix, but each candidate still requires a separate forward pass to generate its output token.
Thus, the decoding cost remains proportional to the number of candidates.
Tree-based speculative decoding~\cite{leviathan:2023, miao:2024, cai:2024} shares
attention computation over a token tree generated by a draft model and discards mispredicted branches.
In contrast, {\tt CD} coalesces identical candidates at runtime, eliminating redundant forward computation beyond attention and KV reuse.

\section{Conclusions}

This paper identified why naive speculation, despite removing step-level synchronization barriers,
can hurt performance in test-time scaling and explored the design space for effective solutions.
We presented {\it \sys}, a serving system that enables efficient speculative execution
through early pruning, computation deduplication, and lazy verification. Through our experiments,
\sys consistently achieved higher throughput and lower end-to-end latency than both the non-speculative
and FastTTS approaches.

\bibliographystyle{ACM-Reference-Format}
\bibliography{references}


\begin{thebibliography}{47}


\ifx \showCODEN    \undefined \def \showCODEN     #1{\unskip}     \fi
\ifx \showDOI      \undefined \def \showDOI       #1{#1}\fi
\ifx \showISBNx    \undefined \def \showISBNx     #1{\unskip}     \fi
\ifx \showISBNxiii \undefined \def \showISBNxiii  #1{\unskip}     \fi
\ifx \showISSN     \undefined \def \showISSN      #1{\unskip}     \fi
\ifx \showLCCN     \undefined \def \showLCCN      #1{\unskip}     \fi
\ifx \shownote     \undefined \def \shownote      #1{#1}          \fi
\ifx \showarticletitle \undefined \def \showarticletitle #1{#1}   \fi
\ifx \showURL      \undefined \def \showURL       {\relax}        \fi
\providecommand\bibfield[2]{#2}
\providecommand\bibinfo[2]{#2}
\providecommand\natexlab[1]{#1}
\providecommand\showeprint[2][]{arXiv:#2}

\bibitem[Agrawal et~al\mbox{.}(2024)]%
        {agrawal:2024}
\bibfield{author}{\bibinfo{person}{Amey Agrawal}, \bibinfo{person}{Nitin
  Kedia}, \bibinfo{person}{Ashish Panwar}, \bibinfo{person}{Jayashree Mohan},
  \bibinfo{person}{Nipun Kwatra}, \bibinfo{person}{Bhargav Gulavani},
  \bibinfo{person}{Alexey Tumanov}, {and} \bibinfo{person}{Ramachandran
  Ramjee}.} \bibinfo{year}{2024}\natexlab{}.
\newblock \showarticletitle{Taming {Throughput-Latency} Tradeoff in {LLM}
  Inference with {Sarathi-Serve}}. In \bibinfo{booktitle}{\emph{18th USENIX
  Symposium on Operating Systems Design and Implementation (OSDI)}}.
\newblock


\bibitem[Ansel et~al\mbox{.}(2024)]%
        {pytorch}
\bibfield{author}{\bibinfo{person}{Jason Ansel}, \bibinfo{person}{Edward Yang},
  \bibinfo{person}{Horace He}, \bibinfo{person}{Natalia Gimelshein},
  \bibinfo{person}{Animesh Jain}, \bibinfo{person}{Michael Voznesensky},
  \bibinfo{person}{Bin Bao}, \bibinfo{person}{Peter Bell},
  \bibinfo{person}{David Berard}, \bibinfo{person}{Evgeni Burovski},
  \bibinfo{person}{Geeta Chauhan}, \bibinfo{person}{Anjali Chourdia},
  \bibinfo{person}{Will Constable}, \bibinfo{person}{Alban Desmaison},
  \bibinfo{person}{Zachary DeVito}, \bibinfo{person}{Elias Ellison},
  \bibinfo{person}{Will Feng}, \bibinfo{person}{Jiong Gong},
  \bibinfo{person}{Michael Gschwind}, \bibinfo{person}{Brian Hirsh},
  \bibinfo{person}{Sherlock Huang}, \bibinfo{person}{Kshiteej Kalambarkar},
  \bibinfo{person}{Laurent Kirsch}, \bibinfo{person}{Michael Lazos},
  \bibinfo{person}{Mario Lezcano}, \bibinfo{person}{Yanbo Liang},
  \bibinfo{person}{Jason Liang}, \bibinfo{person}{Yinghai Lu},
  \bibinfo{person}{CK Luk}, \bibinfo{person}{Bert Maher},
  \bibinfo{person}{Yunjie Pan}, \bibinfo{person}{Christian Puhrsch},
  \bibinfo{person}{Matthias Reso}, \bibinfo{person}{Mark Saroufim},
  \bibinfo{person}{Marcos~Yukio Siraichi}, \bibinfo{person}{Helen Suk},
  \bibinfo{person}{Michael Suo}, \bibinfo{person}{Phil Tillet},
  \bibinfo{person}{Eikan Wang}, \bibinfo{person}{Xiaodong Wang},
  \bibinfo{person}{William Wen}, \bibinfo{person}{Shunting Zhang},
  \bibinfo{person}{Xu Zhao}, \bibinfo{person}{Keren Zhou},
  \bibinfo{person}{Richard Zou}, \bibinfo{person}{Ajit Mathews},
  \bibinfo{person}{Gregory Chanan}, \bibinfo{person}{Peng Wu}, {and}
  \bibinfo{person}{Soumith Chintala}.} \bibinfo{year}{2024}\natexlab{}.
\newblock \showarticletitle{{PyTorch 2: Faster Machine Learning Through Dynamic
  Python Bytecode Transformation and Graph Compilation}}. In
  \bibinfo{booktitle}{\emph{29th ACM International Conference on Architectural
  Support for Programming Languages and Operating Systems (ASPLOS)}}.
\newblock


\bibitem[Beeching et~al\mbox{.}(2024)]%
        {beeching:2024}
\bibfield{author}{\bibinfo{person}{Edward Beeching}, \bibinfo{person}{Lewis
  Tunstall}, {and} \bibinfo{person}{Sasha Rush}.}
  \bibinfo{year}{2024}\natexlab{}.
\newblock \bibinfo{title}{Scaling test-time compute with open models}.
\newblock
\newblock
\newblock
\shownote{\url{https://huggingface.co/spaces/HuggingFaceH4/blogpost-scaling-test-time-compute}
  (Accessed: 2026-04-15)}.


\bibitem[Brown et~al\mbox{.}(2024)]%
        {brown:2024}
\bibfield{author}{\bibinfo{person}{Bradley Brown}, \bibinfo{person}{Jordan
  Juravsky}, \bibinfo{person}{Ryan Ehrlich}, \bibinfo{person}{Ronald Clark},
  \bibinfo{person}{Quoc~V. Le}, \bibinfo{person}{Christopher Ré}, {and}
  \bibinfo{person}{Azalia Mirhoseini}.} \bibinfo{year}{2024}\natexlab{}.
\newblock \showarticletitle{Large Language Monkeys: Scaling Inference Compute
  with Repeated Sampling}.
\newblock  (\bibinfo{year}{2024}).
\newblock
\showeprint[arxiv]{2407.21787}~[cs.LG]
\urldef\tempurl%
\url{https://arxiv.org/abs/2407.21787}
\showURL{%
\tempurl}


\bibitem[Cai et~al\mbox{.}(2024)]%
        {cai:2024}
\bibfield{author}{\bibinfo{person}{Tianle Cai}, \bibinfo{person}{Yuhong Li},
  \bibinfo{person}{Zhengyang Geng}, \bibinfo{person}{Hongwu Peng},
  \bibinfo{person}{Jason~D. Lee}, \bibinfo{person}{Deming Chen}, {and}
  \bibinfo{person}{Tri Dao}.} \bibinfo{year}{2024}\natexlab{}.
\newblock \showarticletitle{MEDUSA: Simple LLM inference acceleration framework
  with multiple decoding heads}. In \bibinfo{booktitle}{\emph{Proceedings of
  the 41st International Conference on Machine Learning (ICML)}}.
\newblock


\bibitem[Cemri et~al\mbox{.}(2025)]%
        {cemri:2025}
\bibfield{author}{\bibinfo{person}{Mert Cemri}, \bibinfo{person}{Nived
  Rajaraman}, \bibinfo{person}{Rishabh Tiwari}, \bibinfo{person}{Xiaoxuan Liu},
  \bibinfo{person}{Kurt Keutzer}, \bibinfo{person}{Ion Stoica},
  \bibinfo{person}{Kannan Ramchandran}, \bibinfo{person}{Ahmad Beirami}, {and}
  \bibinfo{person}{Ziteng Sun}.} \bibinfo{year}{2025}\natexlab{}.
\newblock \bibinfo{title}{$\texttt{SPECS}$: Faster Test-Time Scaling through
  Speculative Drafts}.
\newblock
\newblock
\showeprint[arxiv]{2506.15733}~[cs.AI]
\urldef\tempurl%
\url{https://arxiv.org/abs/2506.15733}
\showURL{%
\tempurl}


\bibitem[Chen et~al\mbox{.}(2026)]%
        {chen:2025}
\bibfield{author}{\bibinfo{person}{Hao~Mark Chen}, \bibinfo{person}{Zhiwen Mo},
  \bibinfo{person}{Guanxi Lu}, \bibinfo{person}{Shuang Liang},
  \bibinfo{person}{Lingxiao Ma}, \bibinfo{person}{Wayne Luk}, {and}
  \bibinfo{person}{Hongxiang Fan}.} \bibinfo{year}{2026}\natexlab{}.
\newblock \showarticletitle{FastTTS: Accelerating Test-Time Scaling for Edge
  LLM Reasoning}. In \bibinfo{booktitle}{\emph{Proceedings of the 31st ACM
  International Conference on Architectural Support for Programming Languages
  and Operating Systems (ASPLOS)}}.
\newblock
\showISBNx{9798400723599}


\bibitem[Cheney(2026)]%
        {fasttts:artifact}
\bibfield{author}{\bibinfo{person}{Mark Cheney}.}
  \bibinfo{year}{2026}\natexlab{}.
\newblock \bibinfo{title}{{FastTTS}: Accelerating Test-Time Scaling for Edge
  {LLM} Reasoning (Artifact)}.
\newblock
  \bibinfo{howpublished}{\url{https://github.com/ihc-fan-lab/FastTTS/blob/1c01d54efe41d9b503895385fdbe08393fa2c513/config.py\#L62-L64}}.
\newblock
\newblock
\shownote{ASPLOS'26 artifact, commit \texttt{1c01d54}}.


\bibitem[Cobbe et~al\mbox{.}(2021)]%
        {gsm}
\bibfield{author}{\bibinfo{person}{Karl Cobbe}, \bibinfo{person}{Vineet
  Kosaraju}, \bibinfo{person}{Mohammad Bavarian}, \bibinfo{person}{Mark Chen},
  \bibinfo{person}{Heewoo Jun}, \bibinfo{person}{Lukasz Kaiser},
  \bibinfo{person}{Matthias Plappert}, \bibinfo{person}{Jerry Tworek},
  \bibinfo{person}{Jacob Hilton}, \bibinfo{person}{Reiichiro Nakano},
  \bibinfo{person}{Christopher Hesse}, {and} \bibinfo{person}{John Schulman}.}
  \bibinfo{year}{2021}\natexlab{}.
\newblock \showarticletitle{Training Verifiers to Solve Math Word Problems}.
\newblock  (\bibinfo{year}{2021}).
\newblock
\showeprint[arxiv]{2110.14168}~[cs.LG]
\urldef\tempurl%
\url{https://arxiv.org/abs/2110.14168}
\showURL{%
\tempurl}


\bibitem[Dao et~al\mbox{.}(2022)]%
        {dao:2022}
\bibfield{author}{\bibinfo{person}{Tri Dao}, \bibinfo{person}{Dan Fu},
  \bibinfo{person}{Stefano Ermon}, \bibinfo{person}{Atri Rudra}, {and}
  \bibinfo{person}{Christopher R\'{e}}.} \bibinfo{year}{2022}\natexlab{}.
\newblock \showarticletitle{FlashAttention: Fast and Memory-Efficient Exact
  Attention with IO-Awareness}. In \bibinfo{booktitle}{\emph{Advances in Neural
  Information Processing Systems (NeurIPS)}}.
\newblock


\bibitem[Gao et~al\mbox{.}(2024)]%
        {gao:2024}
\bibfield{author}{\bibinfo{person}{Bin Gao}, \bibinfo{person}{Zhuomin He},
  \bibinfo{person}{Puru Sharma}, \bibinfo{person}{Qingxuan Kang},
  \bibinfo{person}{Djordje Jevdjic}, \bibinfo{person}{Junbo Deng},
  \bibinfo{person}{Xingkun Yang}, \bibinfo{person}{Zhou Yu}, {and}
  \bibinfo{person}{Pengfei Zuo}.} \bibinfo{year}{2024}\natexlab{}.
\newblock \showarticletitle{{Cost-Efficient} Large Language Model Serving for
  Multi-turn Conversations with {CachedAttention}}. In
  \bibinfo{booktitle}{\emph{2024 USENIX Annual Technical Conference (ATC)}}.
\newblock


\bibitem[Gholami et~al\mbox{.}(2024)]%
        {gholami:2024}
\bibfield{author}{\bibinfo{person}{Amir Gholami}, \bibinfo{person}{Zhewei Yao},
  \bibinfo{person}{Sehoon Kim}, \bibinfo{person}{Coleman Hooper},
  \bibinfo{person}{Michael~W Mahoney}, {and} \bibinfo{person}{Kurt Keutzer}.}
  \bibinfo{year}{2024}\natexlab{}.
\newblock \showarticletitle{Ai and memory wall}.
\newblock \bibinfo{journal}{\emph{IEEE Micro}} \bibinfo{volume}{44},
  \bibinfo{number}{3} (\bibinfo{year}{2024}).
\newblock


\bibitem[He et~al\mbox{.}(2024)]%
        {olympiad}
\bibfield{author}{\bibinfo{person}{Chaoqun He}, \bibinfo{person}{Renjie Luo},
  \bibinfo{person}{Yuzhuo Bai}, \bibinfo{person}{Shengding Hu},
  \bibinfo{person}{Zhen Thai}, \bibinfo{person}{Junhao Shen},
  \bibinfo{person}{Jinyi Hu}, \bibinfo{person}{Xu Han}, \bibinfo{person}{Yujie
  Huang}, \bibinfo{person}{Yuxiang Zhang}, \bibinfo{person}{Jie Liu},
  \bibinfo{person}{Lei Qi}, \bibinfo{person}{Zhiyuan Liu}, {and}
  \bibinfo{person}{Maosong Sun}.} \bibinfo{year}{2024}\natexlab{}.
\newblock \showarticletitle{{O}lympiad{B}ench: A Challenging Benchmark for
  Promoting {AGI} with Olympiad-Level Bilingual Multimodal Scientific
  Problems}. In \bibinfo{booktitle}{\emph{Proceedings of the 62nd Annual
  Meeting of the Association for Computational Linguistics (ACL)}}.
\newblock


\bibitem[Hendrycks et~al\mbox{.}(2021)]%
        {math}
\bibfield{author}{\bibinfo{person}{Dan Hendrycks}, \bibinfo{person}{Collin
  Burns}, \bibinfo{person}{Saurav Kadavath}, \bibinfo{person}{Akul Arora},
  \bibinfo{person}{Steven Basart}, \bibinfo{person}{Eric Tang},
  \bibinfo{person}{Dawn Song}, {and} \bibinfo{person}{Jacob Steinhardt}.}
  \bibinfo{year}{2021}\natexlab{}.
\newblock \showarticletitle{Measuring Mathematical Problem Solving With the
  {MATH} Dataset}. In \bibinfo{booktitle}{\emph{Thirty-fifth Conference on
  Neural Information Processing Systems Datasets and Benchmarks Track (Round
  2)}}.
\newblock


\bibitem[Hinton et~al\mbox{.}(2015)]%
        {hinton2015distillingknowledgeneuralnetwork}
\bibfield{author}{\bibinfo{person}{Geoffrey Hinton}, \bibinfo{person}{Oriol
  Vinyals}, {and} \bibinfo{person}{Jeff Dean}.}
  \bibinfo{year}{2015}\natexlab{}.
\newblock \showarticletitle{Distilling the Knowledge in a Neural Network}.
\newblock \bibinfo{journal}{\emph{arXiv preprint arXiv:1503.02531}}
  (\bibinfo{year}{2015}).
\newblock
\urldef\tempurl%
\url{https://arxiv.org/abs/1503.02531}
\showURL{%
\tempurl}


\bibitem[Hoffmann et~al\mbox{.}(2022)]%
        {hoffmann:2022}
\bibfield{author}{\bibinfo{person}{Jordan Hoffmann}, \bibinfo{person}{Sebastian
  Borgeaud}, \bibinfo{person}{Arthur Mensch}, \bibinfo{person}{Elena
  Buchatskaya}, \bibinfo{person}{Trevor Cai}, \bibinfo{person}{Eliza
  Rutherford}, \bibinfo{person}{Diego de Las~Casas}, \bibinfo{person}{Lisa~Anne
  Hendricks}, \bibinfo{person}{Johannes Welbl}, \bibinfo{person}{Aidan Clark},
  \bibinfo{person}{Tom Hennigan}, \bibinfo{person}{Eric Noland},
  \bibinfo{person}{Katie Millican}, \bibinfo{person}{George van~den Driessche},
  \bibinfo{person}{Bogdan Damoc}, \bibinfo{person}{Aurelia Guy},
  \bibinfo{person}{Simon Osindero}, \bibinfo{person}{Karen Simonyan},
  \bibinfo{person}{Erich Elsen}, \bibinfo{person}{Jack~W. Rae},
  \bibinfo{person}{Oriol Vinyals}, {and} \bibinfo{person}{Laurent Sifre}.}
  \bibinfo{year}{2022}\natexlab{}.
\newblock \showarticletitle{Training Compute-Optimal Large Language Models}.
\newblock  (\bibinfo{year}{2022}).
\newblock
\showeprint[arxiv]{2203.15556}~[cs.CL]
\urldef\tempurl%
\url{https://arxiv.org/abs/2203.15556}
\showURL{%
\tempurl}


\bibitem[Jeong and Ahn(2025)]%
        {jeong:2025}
\bibfield{author}{\bibinfo{person}{Jinwoo Jeong} {and}
  \bibinfo{person}{Jeongseob Ahn}.} \bibinfo{year}{2025}\natexlab{}.
\newblock \showarticletitle{Accelerating LLM Serving for Multi-turn Dialogues
  with Efficient Resource Management}. In \bibinfo{booktitle}{\emph{Proceedings
  of the 30th ACM International Conference on Architectural Support for
  Programming Languages and Operating Systems (ASPLOS)}}.
\newblock


\bibitem[Juravsky et~al\mbox{.}(2024)]%
        {juravsky:2024}
\bibfield{author}{\bibinfo{person}{Jordan Juravsky}, \bibinfo{person}{Bradley
  Brown}, \bibinfo{person}{Ryan Ehrlich}, \bibinfo{person}{Daniel~Y. Fu},
  \bibinfo{person}{Christopher R{\'e}}, {and} \bibinfo{person}{Azalia
  Mirhoseini}.} \bibinfo{year}{2024}\natexlab{}.
\newblock \showarticletitle{Hydragen: High-Throughput {LLM} Inference with
  Shared Prefixes}.
\newblock \bibinfo{journal}{\emph{arXiv preprint arXiv:2402.05099}}
  (\bibinfo{year}{2024}).
\newblock
\showeprint[arxiv]{2402.05099}
\urldef\tempurl%
\url{https://arxiv.org/abs/2402.05099}
\showURL{%
\tempurl}


\bibitem[Kaplan et~al\mbox{.}(2020)]%
        {kaplan:2020}
\bibfield{author}{\bibinfo{person}{Jared Kaplan}, \bibinfo{person}{Sam
  McCandlish}, \bibinfo{person}{Tom Henighan}, \bibinfo{person}{Tom~B. Brown},
  \bibinfo{person}{Benjamin Chess}, \bibinfo{person}{Rewon Child},
  \bibinfo{person}{Scott Gray}, \bibinfo{person}{Alec Radford},
  \bibinfo{person}{Jeffrey Wu}, {and} \bibinfo{person}{Dario Amodei}.}
  \bibinfo{year}{2020}\natexlab{}.
\newblock \showarticletitle{Scaling Laws for Neural Language Models}.
\newblock  (\bibinfo{year}{2020}).
\newblock
\showeprint[arxiv]{2001.08361}~[cs.LG]
\urldef\tempurl%
\url{https://arxiv.org/abs/2001.08361}
\showURL{%
\tempurl}


\bibitem[Kwon et~al\mbox{.}(2023)]%
        {kwon:2023}
\bibfield{author}{\bibinfo{person}{Woosuk Kwon}, \bibinfo{person}{Zhuohan Li},
  \bibinfo{person}{Siyuan Zhuang}, \bibinfo{person}{Ying Sheng},
  \bibinfo{person}{Lianmin Zheng}, \bibinfo{person}{Cody~Hao Yu},
  \bibinfo{person}{Joseph~E. Gonzalez}, \bibinfo{person}{Hao Zhang}, {and}
  \bibinfo{person}{Ion Stoica}.} \bibinfo{year}{2023}\natexlab{}.
\newblock \showarticletitle{Efficient Memory Management for Large Language
  Model Serving with PagedAttention}. In \bibinfo{booktitle}{\emph{29th ACM
  Symposium on Operating Systems Principles (SOSP)}}.
\newblock


\bibitem[Leviathan et~al\mbox{.}(2023)]%
        {leviathan:2023}
\bibfield{author}{\bibinfo{person}{Yaniv Leviathan}, \bibinfo{person}{Matan
  Kalman}, {and} \bibinfo{person}{Yossi Matias}.}
  \bibinfo{year}{2023}\natexlab{}.
\newblock \showarticletitle{Fast inference from transformers via speculative
  decoding}. In \bibinfo{booktitle}{\emph{Proceedings of the 40th International
  Conference on Machine Learning (ICML)}}.
\newblock


\bibitem[Li et~al\mbox{.}(2025a)]%
        {dacheng:2025}
\bibfield{author}{\bibinfo{person}{Dacheng Li}, \bibinfo{person}{Shiyi Cao},
  \bibinfo{person}{Chengkun Cao}, \bibinfo{person}{Xiuyu Li},
  \bibinfo{person}{Shangyin Tan}, \bibinfo{person}{Kurt Keutzer},
  \bibinfo{person}{Jiarong Xing}, \bibinfo{person}{Joseph~E. Gonzalez}, {and}
  \bibinfo{person}{Ion Stoica}.} \bibinfo{year}{2025}\natexlab{a}.
\newblock \showarticletitle{{S}*: Test Time Scaling for Code Generation}. In
  \bibinfo{booktitle}{\emph{Findings of the Association for Computational
  Linguistics (ACL)}}.
\newblock


\bibitem[Li et~al\mbox{.}(2025b)]%
        {sixu:2025}
\bibfield{author}{\bibinfo{person}{Sixu Li}, \bibinfo{person}{Yuzhou Chen},
  \bibinfo{person}{Chaojian Li}, \bibinfo{person}{Yonggan Fu},
  \bibinfo{person}{Zheng Wang}, \bibinfo{person}{Zhongzhi Yu},
  \bibinfo{person}{Haoran You}, \bibinfo{person}{Zhifan Ye},
  \bibinfo{person}{Wei Zhou}, \bibinfo{person}{Yongan Zhang}, {and}
  \bibinfo{person}{Yingyan~(Celine) Lin}.} \bibinfo{year}{2025}\natexlab{b}.
\newblock \showarticletitle{ORCHES: Orchestrated Test-Time-Compute-based LLM
  Reasoning on Collaborative GPU-PIM HEterogeneous System}. In
  \bibinfo{booktitle}{\emph{Proceedings of the 58th IEEE/ACM International
  Symposium on Microarchitecture (MICRO)}}.
\newblock


\bibitem[Lightman et~al\mbox{.}(2024)]%
        {lightman:2024}
\bibfield{author}{\bibinfo{person}{Hunter Lightman}, \bibinfo{person}{Vineet
  Kosaraju}, \bibinfo{person}{Yuri Burda}, \bibinfo{person}{Harrison Edwards},
  \bibinfo{person}{Bowen Baker}, \bibinfo{person}{Teddy Lee},
  \bibinfo{person}{Jan Leike}, \bibinfo{person}{John Schulman},
  \bibinfo{person}{Ilya Sutskever}, {and} \bibinfo{person}{Karl Cobbe}.}
  \bibinfo{year}{2024}\natexlab{}.
\newblock \showarticletitle{Let's Verify Step by Step}. In
  \bibinfo{booktitle}{\emph{The Twelfth International Conference on Learning
  Representations (ICLR)}}.
\newblock


\bibitem[Madaan et~al\mbox{.}(2023)]%
        {madaan:2023}
\bibfield{author}{\bibinfo{person}{Aman Madaan}, \bibinfo{person}{Niket
  Tandon}, \bibinfo{person}{Prakhar Gupta}, \bibinfo{person}{Skyler Hallinan},
  \bibinfo{person}{Luyu Gao}, \bibinfo{person}{Sarah Wiegreffe},
  \bibinfo{person}{Uri Alon}, \bibinfo{person}{Nouha Dziri},
  \bibinfo{person}{Shrimai Prabhumoye}, \bibinfo{person}{Yiming Yang},
  \bibinfo{person}{Shashank Gupta}, \bibinfo{person}{Bodhisattwa~Prasad
  Majumder}, \bibinfo{person}{Katherine Hermann}, \bibinfo{person}{Sean
  Welleck}, \bibinfo{person}{Amir Yazdanbakhsh}, {and} \bibinfo{person}{Peter
  Clark}.} \bibinfo{year}{2023}\natexlab{}.
\newblock \showarticletitle{Self-Refine: Iterative Refinement with
  Self-Feedback}. In \bibinfo{booktitle}{\emph{Advances in Neural Information
  Processing Systems (NeurIPS)}}.
\newblock


\bibitem[Meta(2023)]%
        {touvron:2023}
\bibfield{author}{\bibinfo{person}{Llama Team AI~@ Meta}.}
  \bibinfo{year}{2023}\natexlab{}.
\newblock \showarticletitle{Llama 2: Open Foundation and Fine-Tuned Chat
  Models}.
\newblock  (\bibinfo{year}{2023}).
\newblock
\showeprint[arxiv]{2307.09288}~[cs.CL]
\urldef\tempurl%
\url{https://arxiv.org/abs/2307.09288}
\showURL{%
\tempurl}


\bibitem[Miao et~al\mbox{.}(2024)]%
        {miao:2024}
\bibfield{author}{\bibinfo{person}{Xupeng Miao}, \bibinfo{person}{Gabriele
  Oliaro}, \bibinfo{person}{Zhihao Zhang}, \bibinfo{person}{Xinhao Cheng},
  \bibinfo{person}{Zeyu Wang}, \bibinfo{person}{Zhengxin Zhang},
  \bibinfo{person}{Rae Ying~Yee Wong}, \bibinfo{person}{Alan Zhu},
  \bibinfo{person}{Lijie Yang}, \bibinfo{person}{Xiaoxiang Shi},
  \bibinfo{person}{Chunan Shi}, \bibinfo{person}{Zhuoming Chen},
  \bibinfo{person}{Daiyaan Arfeen}, \bibinfo{person}{Reyna Abhyankar}, {and}
  \bibinfo{person}{Zhihao Jia}.} \bibinfo{year}{2024}\natexlab{}.
\newblock \showarticletitle{SpecInfer: Accelerating Large Language Model
  Serving with Tree-based Speculative Inference and Verification}. In
  \bibinfo{booktitle}{\emph{Proceedings of the 29th ACM International
  Conference on Architectural Support for Programming Languages and Operating
  Systems (ASPLOS)}}.
\newblock


\bibitem[Muennighoff et~al\mbox{.}(2025)]%
        {muennighoff:2025}
\bibfield{author}{\bibinfo{person}{Niklas Muennighoff}, \bibinfo{person}{Zitong
  Yang}, \bibinfo{person}{Weijia Shi}, \bibinfo{person}{Xiang~Lisa Li},
  \bibinfo{person}{Li Fei-Fei}, \bibinfo{person}{Hannaneh Hajishirzi},
  \bibinfo{person}{Luke Zettlemoyer}, \bibinfo{person}{Percy Liang},
  \bibinfo{person}{Emmanuel Candes}, {and} \bibinfo{person}{Tatsunori
  Hashimoto}.} \bibinfo{year}{2025}\natexlab{}.
\newblock \showarticletitle{s1: Simple test-time scaling}. In
  \bibinfo{booktitle}{\emph{Proceedings of the 2025 Conference on Empirical
  Methods in Natural Language Processing (EMNLP)}}.
\newblock


\bibitem[NVIDIA et~al\mbox{.}(2025)]%
        {cuda}
\bibfield{author}{\bibinfo{person}{NVIDIA}, \bibinfo{person}{Péter
  Vingelmann}, {and} \bibinfo{person}{Frank~H.P. Fitzek}.}
  \bibinfo{year}{2025}\natexlab{}.
\newblock \bibinfo{title}{{CUDA}, release: 12.8}.
\newblock
\newblock
\newblock
\shownote{\url{https://developer.nvidia.com/cuda-12-8-0-download-archive}
  (Accessed: 2026-04-15)}.


\bibitem[{OpenAI}(2024)]%
        {openai:2024}
\bibfield{author}{\bibinfo{person}{{OpenAI}}.} \bibinfo{year}{2024}\natexlab{}.
\newblock \bibinfo{title}{Learning to Reason with LLMs}.
\newblock
\newblock
\newblock
\shownote{\url{https://openai.com/index/learning-to-reason-with-llms/}
  (Accessed: 2026-04-15)}.


\bibitem[Snell et~al\mbox{.}(2025)]%
        {snell:2025}
\bibfield{author}{\bibinfo{person}{Charlie~Victor Snell},
  \bibinfo{person}{Jaehoon Lee}, \bibinfo{person}{Kelvin Xu}, {and}
  \bibinfo{person}{Aviral Kumar}.} \bibinfo{year}{2025}\natexlab{}.
\newblock \showarticletitle{Scaling {LLM} Test-Time Compute Optimally Can be
  More Effective than Scaling Parameters for Reasoning}. In
  \bibinfo{booktitle}{\emph{The Thirteenth International Conference on Learning
  Representations (ICLR)}}.
\newblock


\bibitem[Team(2025)]%
        {deepseekai:2025}
\bibfield{author}{\bibinfo{person}{DeepSeek-AI Team}.}
  \bibinfo{year}{2025}\natexlab{}.
\newblock \bibinfo{title}{DeepSeek-R1: Incentivizing Reasoning Capability in
  LLMs via Reinforcement Learning}.
\newblock
\newblock
\showeprint[arxiv]{2501.12948}~[cs.CL]
\urldef\tempurl%
\url{https://arxiv.org/abs/2501.12948}
\showURL{%
\tempurl}


\bibitem[Team(2024a)]%
        {achiam:2024}
\bibfield{author}{\bibinfo{person}{OpenAI Team}.}
  \bibinfo{year}{2024}\natexlab{a}.
\newblock \bibinfo{title}{GPT-4 Technical Report}.
\newblock
\newblock
\showeprint[arxiv]{2303.08774}~[cs.CL]
\urldef\tempurl%
\url{https://arxiv.org/abs/2303.08774}
\showURL{%
\tempurl}


\bibitem[Team(2024b)]%
        {hurst:2024}
\bibfield{author}{\bibinfo{person}{OpenAI Team}.}
  \bibinfo{year}{2024}\natexlab{b}.
\newblock \bibinfo{title}{GPT-4o System Card}.
\newblock
\newblock
\showeprint[arxiv]{2410.21276}~[cs.CL]
\urldef\tempurl%
\url{https://arxiv.org/abs/2410.21276}
\showURL{%
\tempurl}


\bibitem[Uesato et~al\mbox{.}(2022)]%
        {uesato:2022}
\bibfield{author}{\bibinfo{person}{Jonathan Uesato}, \bibinfo{person}{Nate
  Kushman}, \bibinfo{person}{Ramana Kumar}, \bibinfo{person}{Francis Song},
  \bibinfo{person}{Noah Siegel}, \bibinfo{person}{Lisa Wang},
  \bibinfo{person}{Antonia Creswell}, \bibinfo{person}{Geoffrey Irving}, {and}
  \bibinfo{person}{Irina Higgins}.} \bibinfo{year}{2022}\natexlab{}.
\newblock \showarticletitle{Solving math word problems with process- and
  outcome-based feedback}.
\newblock  (\bibinfo{year}{2022}).
\newblock
\showeprint[arxiv]{2211.14275}~[cs.LG]
\urldef\tempurl%
\url{https://arxiv.org/abs/2211.14275}
\showURL{%
\tempurl}


\bibitem[Vaswani et~al\mbox{.}(2017)]%
        {vaswani:2017}
\bibfield{author}{\bibinfo{person}{Ashish Vaswani}, \bibinfo{person}{Noam
  Shazeer}, \bibinfo{person}{Niki Parmar}, \bibinfo{person}{Jakob Uszkoreit},
  \bibinfo{person}{Llion Jones}, \bibinfo{person}{Aidan~N Gomez},
  \bibinfo{person}{\L~ukasz Kaiser}, {and} \bibinfo{person}{Illia
  P~olosukhin}.} \bibinfo{year}{2017}\natexlab{}.
\newblock \showarticletitle{Attention is All you Need}. In
  \bibinfo{booktitle}{\emph{Advances in Neural Information Processing Systems
  (NeurIPS)}}.
\newblock


\bibitem[Wan et~al\mbox{.}(2024)]%
        {ziyu:2024}
\bibfield{author}{\bibinfo{person}{Ziyu Wan}, \bibinfo{person}{Xidong Feng},
  \bibinfo{person}{Muning Wen}, \bibinfo{person}{Stephen~Marcus McAleer},
  \bibinfo{person}{Ying Wen}, \bibinfo{person}{Weinan Zhang}, {and}
  \bibinfo{person}{Jun Wang}.} \bibinfo{year}{2024}\natexlab{}.
\newblock \showarticletitle{AlphaZero-like tree-search can guide large language
  model decoding and training}. In \bibinfo{booktitle}{\emph{Proceedings of the
  41st International Conference on Machine Learning (ICML)}}.
\newblock


\bibitem[Wang et~al\mbox{.}(2024)]%
        {wang:2024}
\bibfield{author}{\bibinfo{person}{Jun Wang}, \bibinfo{person}{Meng Fang},
  \bibinfo{person}{Ziyu Wan}, \bibinfo{person}{Muning Wen},
  \bibinfo{person}{Jiachen Zhu}, \bibinfo{person}{Anjie Liu},
  \bibinfo{person}{Ziqin Gong}, \bibinfo{person}{Yan Song},
  \bibinfo{person}{Lei Chen}, \bibinfo{person}{Lionel~M. Ni},
  \bibinfo{person}{Linyi Yang}, \bibinfo{person}{Ying Wen}, {and}
  \bibinfo{person}{Weinan Zhang}.} \bibinfo{year}{2024}\natexlab{}.
\newblock \showarticletitle{OpenR: An Open Source Framework for Advanced
  Reasoning with Large Language Models}.
\newblock \bibinfo{journal}{\emph{arXiv preprint arXiv:2410.09671}}
  (\bibinfo{year}{2024}).
\newblock
\showeprint[arxiv]{2410.09671}~[cs.AI]
\urldef\tempurl%
\url{https://arxiv.org/abs/2410.09671}
\showURL{%
\tempurl}


\bibitem[Wu et~al\mbox{.}(2025)]%
        {wu:2025}
\bibfield{author}{\bibinfo{person}{Yangzhen Wu}, \bibinfo{person}{Zhiqing Sun},
  \bibinfo{person}{Shanda Li}, \bibinfo{person}{Sean Welleck}, {and}
  \bibinfo{person}{Yiming Yang}.} \bibinfo{year}{2025}\natexlab{}.
\newblock \showarticletitle{Inference Scaling Laws: An Empirical Analysis of
  Compute-Optimal Inference for Problem-Solving with Language Models}. In
  \bibinfo{booktitle}{\emph{The Thirteenth International Conference on Learning
  Representations (ICLR)}}.
\newblock


\bibitem[Yao et~al\mbox{.}(2025)]%
        {yao:2025}
\bibfield{author}{\bibinfo{person}{Jiayi Yao}, \bibinfo{person}{Hanchen Li},
  \bibinfo{person}{Yuhan Liu}, \bibinfo{person}{Siddhant Ray},
  \bibinfo{person}{Yihua Cheng}, \bibinfo{person}{Qizheng Zhang},
  \bibinfo{person}{Kuntai Du}, \bibinfo{person}{Shan Lu}, {and}
  \bibinfo{person}{Junchen Jiang}.} \bibinfo{year}{2025}\natexlab{}.
\newblock \showarticletitle{CacheBlend: Fast Large Language Model Serving for
  RAG with Cached Knowledge Fusion}. In \bibinfo{booktitle}{\emph{Proceedings
  of the Twentieth European Conference on Computer Systems (EuroSys)}}.
\newblock


\bibitem[Yao et~al\mbox{.}(2023)]%
        {shunyu:2023}
\bibfield{author}{\bibinfo{person}{Shunyu Yao}, \bibinfo{person}{Dian Yu},
  \bibinfo{person}{Jeffrey Zhao}, \bibinfo{person}{Izhak Shafran},
  \bibinfo{person}{Thomas~L. Griffiths}, \bibinfo{person}{Yuan Cao}, {and}
  \bibinfo{person}{Karthik Narasimhan}.} \bibinfo{year}{2023}\natexlab{}.
\newblock \showarticletitle{Tree of thoughts: deliberate problem solving with
  large language models}. In \bibinfo{booktitle}{\emph{Proceedings of the 37th
  International Conference on Neural Information Processing Systems
  (NeurIPS)}}.
\newblock


\bibitem[Ye et~al\mbox{.}(2025)]%
        {ye:2025}
\bibfield{author}{\bibinfo{person}{Zihao Ye}, \bibinfo{person}{Lequn Chen},
  \bibinfo{person}{Ruihang Lai}, \bibinfo{person}{Wuwei Lin},
  \bibinfo{person}{Yineng Zhang}, \bibinfo{person}{Stephanie Wang},
  \bibinfo{person}{Tianqi Chen}, \bibinfo{person}{Baris Kasikci},
  \bibinfo{person}{Vinod Grover}, \bibinfo{person}{Arvind Krishnamurthy}, {and}
  \bibinfo{person}{Luis Ceze}.} \bibinfo{year}{2025}\natexlab{}.
\newblock \showarticletitle{FlashInfer: Efficient and Customizable Attention
  Engine for {LLM} Inference Serving}. In \bibinfo{booktitle}{\emph{Eighth
  Conference on Machine Learning and Systems (MLSys)}}.
\newblock


\bibitem[Yi et~al\mbox{.}(2026)]%
        {yi:2026}
\bibfield{author}{\bibinfo{person}{Jinjun Yi}, \bibinfo{person}{Zhixin Zhao},
  \bibinfo{person}{Yitao Hu}, \bibinfo{person}{Ke Yan}, \bibinfo{person}{Weiwei
  Sun}, \bibinfo{person}{Hao Wang}, \bibinfo{person}{Laiping Zhao},
  \bibinfo{person}{Yuhao Zhang}, \bibinfo{person}{Wenxin Li}, {and}
  \bibinfo{person}{Keqiu Li}.} \bibinfo{year}{2026}\natexlab{}.
\newblock \showarticletitle{PAT: Accelerating LLM Decoding via Prefix-Aware
  Attention with Resource Efficient Multi-Tile Kernel}. In
  \bibinfo{booktitle}{\emph{Proceedings of the 31st ACM International
  Conference on Architectural Support for Programming Languages and Operating
  Systems (ASPLOS)}}.
\newblock


\bibitem[Yu et~al\mbox{.}(2022)]%
        {yu:2022}
\bibfield{author}{\bibinfo{person}{Gyeong-In Yu}, \bibinfo{person}{Joo~Seong
  Jeong}, \bibinfo{person}{Geon-Woo Kim}, \bibinfo{person}{Soojeong Kim}, {and}
  \bibinfo{person}{Byung-Gon Chun}.} \bibinfo{year}{2022}\natexlab{}.
\newblock \showarticletitle{Orca: A Distributed Serving System for
  {Transformer-Based} Generative Models}. In \bibinfo{booktitle}{\emph{16th
  USENIX Symposium on Operating Systems Design and Implementation (OSDI)}}.
\newblock
\showISBNx{978-1-939133-28-1}


\bibitem[Zhang et~al\mbox{.}(2025)]%
        {qwen-prm}
\bibfield{author}{\bibinfo{person}{Zhenru Zhang}, \bibinfo{person}{Chujie
  Zheng}, \bibinfo{person}{Yangzhen Wu}, \bibinfo{person}{Beichen Zhang},
  \bibinfo{person}{Runji Lin}, \bibinfo{person}{Bowen Yu},
  \bibinfo{person}{Dayiheng Liu}, \bibinfo{person}{Jingren Zhou}, {and}
  \bibinfo{person}{Junyang Lin}.} \bibinfo{year}{2025}\natexlab{}.
\newblock \showarticletitle{The Lessons of Developing Process Reward Models in
  Mathematical Reasoning}. In \bibinfo{booktitle}{\emph{Findings of the
  Association for Computational Linguistics (ACL)}}.
\newblock


\bibitem[Zheng et~al\mbox{.}(2024)]%
        {zheng:2024}
\bibfield{author}{\bibinfo{person}{Lianmin Zheng}, \bibinfo{person}{Liangsheng
  Yin}, \bibinfo{person}{Zhiqiang Xie}, \bibinfo{person}{Chuyue Sun},
  \bibinfo{person}{Jeff Huang}, \bibinfo{person}{Cody~Hao Yu},
  \bibinfo{person}{Shiyi Cao}, \bibinfo{person}{Christos Kozyrakis},
  \bibinfo{person}{Ion Stoica}, \bibinfo{person}{Joseph~E. Gonzalez},
  \bibinfo{person}{Clark Barrett}, {and} \bibinfo{person}{Ying Sheng}.}
  \bibinfo{year}{2024}\natexlab{}.
\newblock \showarticletitle{SGLang: efficient execution of structured language
  model programs}. In \bibinfo{booktitle}{\emph{Advances in Neural Information
  Processing Systems (NeurIPS)}}.
\newblock


\bibitem[Zhong et~al\mbox{.}(2026)]%
        {zhong:2026}
\bibfield{author}{\bibinfo{person}{Shuzhang Zhong}, \bibinfo{person}{Haochen
  Huang}, \bibinfo{person}{Shengxuan Qiu}, \bibinfo{person}{Pengfei Zuo},
  \bibinfo{person}{Runsheng Wang}, {and} \bibinfo{person}{Meng Li}.}
  \bibinfo{year}{2026}\natexlab{}.
\newblock \showarticletitle{Breaking the Reward Barrier: Accelerating
  {Tree-of-Thought} Reasoning via Speculative Exploration}. In
  \bibinfo{booktitle}{\emph{20th USENIX Symposium on Operating Systems Design
  and Implementation (OSDI)}}.
\newblock


\end{thebibliography}
\end{document}